\documentclass[twocolumn]{aastex701}

\usepackage{etoolbox}
\makeatletter
\let\original@label\label
\makeatother

\usepackage{multirow}
\usepackage{subfigure}
\usepackage{siunitx}

\usepackage{dcolumn}

\usepackage{amsmath}
\usepackage{enumitem}
\usepackage{todonotes}
\usepackage{makecell}
\usepackage{comment}
\usepackage{gensymb}

\usepackage{hyperref}

\hypersetup{
    colorlinks=true,
    linkcolor=blue,
    filecolor=magenta,      
    urlcolor=cyan,
    pdftitle={Improving Pulsar Timing Precision with Single Pulse Fluence Clustering},
    pdfpagemode=FullScreen,
    }

\begin{document}

\title{The NANOGrav 15-Year Data Set: Improved Timing Precision With VLBI Astrometric Priors}

\author[0000-0002-5176-2924]{Sofia V. Sosa Fiscella}
\affiliation{School of Physics and Astronomy, Rochester Institute of Technology, Rochester, NY 14623, USA}
\affiliation{Laboratory for Multiwavelength Astrophysics, Rochester Institute of Technology, Rochester, NY 14623, USA}
\email{sophia.sosa@nanograv.org}
\author[0000-0003-0721-651X]{Michael T. Lam}
\affiliation{SETI Institute, 339 N Bernardo Ave Suite 200, Mountain View, CA 94043, USA}
\affiliation{School of Physics and Astronomy, Rochester Institute of Technology, Rochester, NY 14623, USA}
\affiliation{Laboratory for Multiwavelength Astrophysics, Rochester Institute of Technology, Rochester, NY 14623, USA}
\email{michael.lam@nanograv.org}

\author[0000-0001-5134-3925]{Gabriella Agazie}
\affiliation{Center for Gravitation, Cosmology and Astrophysics, Department of Physics and Astronomy, University of Wisconsin-Milwaukee,\\ P.O. Box 413, Milwaukee, WI 53201, USA}
\email{gabriella.agazie@nanograv.org}
\author[0000-0002-8935-9882]{Akash Anumarlapudi}
\affiliation{Department of Physics and Astronomy, University of North Carolina, Chapel Hill, NC 27599, USA}
\email{akasha@unc.edu}
\author[0000-0003-0638-3340]{Anne M. Archibald}
\affiliation{Newcastle University, NE1 7RU, UK}
\email{anne.archibald@nanograv.org}
\author[0009-0008-6187-8753]{Zaven Arzoumanian}
\affiliation{X-Ray Astrophysics Laboratory, NASA Goddard Space Flight Center, Code 662, Greenbelt, MD 20771, USA}
\email{zaven.arzoumanian@nanograv.org}
\author[0000-0003-2745-753X]{Paul T. Baker}
\affiliation{Department of Physics and Astronomy, Widener University, One University Place, Chester, PA 19013, USA}
\email{paul.baker@nanograv.org}
\author[0000-0003-3053-6538]{Paul R. Brook}
\affiliation{Institute for Gravitational Wave Astronomy and School of Physics and Astronomy, University of Birmingham, Edgbaston, Birmingham B15 2TT, UK}
\email{paul.brook@nanograv.org}
\author[0000-0002-6039-692X]{H. Thankful Cromartie}
\affiliation{National Research Council Research Associate, National Academy of Sciences, Washington, DC 20001, USA resident at Naval Research Laboratory, Washington, DC 20375, USA}
\email{thankful.cromartie@nanograv.org}
\author[0000-0002-1529-5169]{Kathryn Crowter}
\affiliation{Department of Physics and Astronomy, University of British Columbia, 6224 Agricultural Road, Vancouver, BC V6T 1Z1, Canada}
\email{kathryn.crowter@nanograv.org}

\author[0009-0007-6217-9236]{Mar\'ia Silvina De Biasi}
\affiliation{Facultad de Ciencias Astron\'omicas y Geof\'isicas, Universidad Nacional de La Plata, Paseo del Bosque, B1900FWA La Plata, Argentina}
\affiliation{Instituto de Astrofísica de La Plata, Consejo Nacional de Investigaciones Científicas y Técnicas, Paseo del Bosque, B1900FWA La Plata, Argentina}
\email{debiasi@fcaglp.unlp.edu.ar}

\author[0000-0002-2185-1790]{Megan E. DeCesar}
\altaffiliation{Resident at the Naval Research Laboratory}
\affiliation{Department of Physics and Astronomy, George Mason University, Fairfax, VA 22030, resident at the U.S. Naval Research Laboratory, Washington, DC 20375, USA}
\email{megan.decesar@nanograv.org}
\author[0000-0002-6664-965X]{Paul B. Demorest}
\affiliation{National Radio Astronomy Observatory, 1003 Lopezville Rd., Socorro, NM 87801, USA}
\email{paul.demorest@nanograv.org}
\author[0000-0001-8885-6388]{Timothy Dolch}
\affiliation{Department of Physics, Hillsdale College, 33 E. College Street, Hillsdale, MI 49242, USA}
\affiliation{Eureka Scientific, 2452 Delmer Street, Suite 100, Oakland, CA 94602-3017, USA}
\email{timothy.dolch@nanograv.org}
\author[0000-0001-7828-7708]{Elizabeth C. Ferrara}
\affiliation{Department of Astronomy, University of Maryland, College Park, MD 20742, USA}
\affiliation{Center for Research and Exploration in Space Science and Technology, NASA/GSFC, Greenbelt, MD 20771}
\affiliation{NASA Goddard Space Flight Center, Greenbelt, MD 20771, USA}
\email{elizabeth.ferrara@nanograv.org}
\author[0000-0001-5645-5336]{William Fiore}
\affiliation{Department of Physics and Astronomy, University of British Columbia, 6224 Agricultural Road, Vancouver, BC V6T 1Z1, Canada}
\email{william.fiore@nanograv.org}
\author[0000-0001-8384-5049]{Emmanuel Fonseca}
\affiliation{Department of Physics and Astronomy, West Virginia University, P.O. Box 6315, Morgantown, WV 26506, USA}
\affiliation{Center for Gravitational Waves and Cosmology, West Virginia University, Chestnut Ridge Research Building, Morgantown, WV 26505, USA}
\email{emmanuel.fonseca@nanograv.org}
\author[0000-0001-7624-4616]{Gabriel E. Freedman}
\affiliation{Center for Gravitation, Cosmology and Astrophysics, Department of Physics and Astronomy, University of Wisconsin-Milwaukee,\\ P.O. Box 413, Milwaukee, WI 53201, USA}
\email{gabriel.freedman@nanograv.org}
\author[0000-0001-6166-9646]{Nate Garver-Daniels}
\affiliation{Department of Physics and Astronomy, West Virginia University, P.O. Box 6315, Morgantown, WV 26506, USA}
\affiliation{Center for Gravitational Waves and Cosmology, West Virginia University, Chestnut Ridge Research Building, Morgantown, WV 26505, USA}
\email{nathaniel.garver-daniels@nanograv.org}
\author[0000-0001-8158-683X]{Peter A. Gentile}
\affiliation{Department of Physics and Astronomy, West Virginia University, P.O. Box 6315, Morgantown, WV 26506, USA}
\affiliation{Center for Gravitational Waves and Cosmology, West Virginia University, Chestnut Ridge Research Building, Morgantown, WV 26505, USA}
\email{peter.gentile@nanograv.org}
\author[0000-0003-4090-9780]{Joseph Glaser}
\affiliation{Department of Physics and Astronomy, West Virginia University, P.O. Box 6315, Morgantown, WV 26506, USA}
\affiliation{Center for Gravitational Waves and Cosmology, West Virginia University, Chestnut Ridge Research Building, Morgantown, WV 26505, USA}
\email{joseph.glaser@nanograv.org}
\author[0000-0003-1884-348X]{Deborah C. Good}
\affiliation{Department of Physics and Astronomy, University of Montana, 32 Campus Drive, Missoula, MT 59812}
\email{deborah.good@nanograv.org}
\author[0000-0003-2742-3321]{Jeffrey S. Hazboun}
\affiliation{Department of Physics, Oregon State University, Corvallis, OR 97331, USA}
\email{jeffrey.hazboun@nanograv.org}
\author[0000-0003-1082-2342]{Ross J. Jennings}
\altaffiliation{NANOGrav Physics Frontiers Center Postdoctoral Fellow}
\affiliation{Department of Physics and Astronomy, West Virginia University, P.O. Box 6315, Morgantown, WV 26506, USA}
\affiliation{Center for Gravitational Waves and Cosmology, West Virginia University, Chestnut Ridge Research Building, Morgantown, WV 26505, USA}
\email{ross.jennings@nanograv.org}
\author[0000-0001-6607-3710]{Megan L. Jones}
\affiliation{Center for Gravitation, Cosmology and Astrophysics, Department of Physics and Astronomy, University of Wisconsin-Milwaukee,\\ P.O. Box 413, Milwaukee, WI 53201, USA}
\email{megan.jones@nanograv.org}
\author[0000-0001-6295-2881]{David L. Kaplan}
\affiliation{Center for Gravitation, Cosmology and Astrophysics, Department of Physics and Astronomy, University of Wisconsin-Milwaukee,\\ P.O. Box 413, Milwaukee, WI 53201, USA}
\email{kaplan@uwm.edu}
\author[0000-0002-0893-4073]{Matthew Kerr}
\affiliation{Space Science Division, Naval Research Laboratory, Washington, DC 20375-5352, USA}
\email{matthew.kerr@nanograv.org}
\author[0000-0003-1301-966X]{Duncan R. Lorimer}
\affiliation{Department of Physics and Astronomy, West Virginia University, P.O. Box 6315, Morgantown, WV 26506, USA}
\affiliation{Center for Gravitational Waves and Cosmology, West Virginia University, Chestnut Ridge Research Building, Morgantown, WV 26505, USA}
\email{duncan.lorimer@nanograv.org}
\author[0000-0001-5373-5914]{Jing Luo}
\altaffiliation{Deceased}
\affiliation{Department of Astronomy \& Astrophysics, University of Toronto, 50 Saint George Street, Toronto, ON M5S 3H4, Canada}
\email{jing.luo@nanograv.org}
\author[0000-0001-5229-7430]{Ryan S. Lynch}
\affiliation{Green Bank Observatory, P.O. Box 2, Green Bank, WV 24944, USA}
\email{ryan.lynch@nanograv.org}
\author[0000-0001-5481-7559]{Alexander McEwen}
\affiliation{Center for Gravitation, Cosmology and Astrophysics, Department of Physics and Astronomy, University of Wisconsin-Milwaukee,\\ P.O. Box 413, Milwaukee, WI 53201, USA}
\email{alexander.mcewen@nanograv.org}
\author[0000-0001-7697-7422]{Maura A. McLaughlin}
\affiliation{Department of Physics and Astronomy, West Virginia University, P.O. Box 6315, Morgantown, WV 26506, USA}
\affiliation{Center for Gravitational Waves and Cosmology, West Virginia University, Chestnut Ridge Research Building, Morgantown, WV 26505, USA}
\email{maura.mclaughlin@nanograv.org}
\author[0000-0002-4642-1260]{Natasha McMann}
\affiliation{Department of Physics and Astronomy, Vanderbilt University, 2301 Vanderbilt Place, Nashville, TN 37235, USA}
\email{natasha.mcmann@nanograv.org}
\author[0000-0001-8845-1225]{Bradley W. Meyers}
\affiliation{Australian SKA Regional Centre (AusSRC), Curtin University, Bentley, WA 6102, Australia}
\affiliation{International Centre for Radio Astronomy Research (ICRAR), Curtin University, Bentley, WA 6102, Australia}
\email{bradley.meyers@nanograv.org}
\author[0000-0002-3616-5160]{Cherry Ng}
\affiliation{Dunlap Institute for Astronomy and Astrophysics, University of Toronto, 50 St. George St., Toronto, ON M5S 3H4, Canada}
\email{cherry.ng@nanograv.org}
\author[0000-0002-6709-2566]{David J. Nice}
\affiliation{Department of Physics, Lafayette College, Easton, PA 18042, USA}
\email{niced@lafayette.edu}
\author[0000-0001-5465-2889]{Timothy T. Pennucci}
\affiliation{Institute of Physics and Astronomy, E\"{o}tv\"{o}s Lor\'{a}nd University, P\'{a}zm\'{a}ny P. s. 1/A, 1117 Budapest, Hungary}
\email{timothy.pennucci@nanograv.org}
\author[0000-0002-8509-5947]{Benetge B. P. Perera}
\affiliation{Arecibo Observatory, HC3 Box 53995, Arecibo, PR 00612, USA}
\email{benetge.perera@nanograv.org}
\author[0000-0002-8826-1285]{Nihan S. Pol}
\affiliation{Department of Physics, Texas Tech University, Box 41051, Lubbock, TX 79409, USA}
\email{nihan.pol@nanograv.org}
\author[0000-0002-2074-4360]{Henri A. Radovan}
\affiliation{Department of Physics, University of Puerto Rico, Mayag\"{u}ez, PR 00681, USA}
\email{henri.radovan@nanograv.org}
\author[0000-0001-5799-9714]{Scott M. Ransom}
\affiliation{National Radio Astronomy Observatory, 520 Edgemont Road, Charlottesville, VA 22903, USA}
\email{sransom@nrao.edu}
\author[0000-0002-5297-5278]{Paul S. Ray}
\affiliation{Space Science Division, Naval Research Laboratory, Washington, DC 20375-5352, USA}
\email{paul.ray@nanograv.org}
\author[0000-0003-4391-936X]{Ann Schmiedekamp}
\affiliation{Department of Physics, Penn State Abington, Abington, PA 19001, USA}
\email{ann.schmiedekamp@nanograv.org}
\author[0000-0002-1283-2184]{Carl Schmiedekamp}
\affiliation{Department of Physics, Penn State Abington, Abington, PA 19001, USA}
\email{carl.schmiedekamp@nanograv.org}
\author[0000-0002-7283-1124]{Brent J. Shapiro-Albert}
\affiliation{Department of Physics and Astronomy, West Virginia University, P.O. Box 6315, Morgantown, WV 26506, USA}
\affiliation{Center for Gravitational Waves and Cosmology, West Virginia University, Chestnut Ridge Research Building, Morgantown, WV 26505, USA}
\affiliation{Giant Army, 915A 17th Ave, Seattle WA 98122}
\email{brent.shapiro-albert@nanograv.org}
\author[0000-0001-9784-8670]{Ingrid H. Stairs}
\affiliation{Department of Physics and Astronomy, University of British Columbia, 6224 Agricultural Road, Vancouver, BC V6T 1Z1, Canada}
\email{stairs@astro.ubc.ca}
\author[0000-0002-7261-594X]{Kevin Stovall}
\affiliation{National Radio Astronomy Observatory, 1003 Lopezville Rd., Socorro, NM 87801, USA}
\email{kevin.stovall@nanograv.org}
\author[0000-0002-2820-0931]{Abhimanyu Susobhanan}
\affiliation{Max-Planck-Institut f{\"u}r Gravitationsphysik (Albert-Einstein-Institut), Callinstra{\ss}e 38, D-30167 Hannover, Germany\\}
\email{abhimanyu.susobhanan@nanograv.org}
\author[0000-0002-1075-3837]{Joseph K. Swiggum}
\altaffiliation{NANOGrav Physics Frontiers Center Postdoctoral Fellow}
\affiliation{Department of Physics, Lafayette College, Easton, PA 18042, USA}
\email{joseph.swiggum@nanograv.org}
\author[0000-0001-9678-0299]{Haley M. Wahl}
\affiliation{Department of Physics and Astronomy, West Virginia University, P.O. Box 6315, Morgantown, WV 26506, USA}
\affiliation{Center for Gravitational Waves and Cosmology, West Virginia University, Chestnut Ridge Research Building, Morgantown, WV 26505, USA}
\email{haley.wahl@nanograv.org}




\begin{abstract}
Accurate pulsar astrometric estimates play an essential role in almost all high-precision pulsar timing experiments. Traditional pulsar timing techniques refine these estimates by including them as free parameters when fitting a model to observed pulse time-of-arrival measurements. However, reliable sub-milliarcsecond astrometric estimations require years of observations and, even then, power from red noise can be inadvertently absorbed into astrometric parameter fits, biasing the resulting estimations and reducing our sensitivity to red noise processes, including gravitational waves (GWs). In this work, we seek to mitigate these shortcomings by using pulsar astrometric estimates derived from Very Long Baseline Interferometry (VLBI) as priors for the timing fit. First, we calibrated a tie between the International Celestial Reference Frame, as realized by the Radio Fundamental Catalog, and the JPL's DE440 planetary ephemerides frame with a precision of $\sim0.6$ mas. Then, we used the VLBI-informed priors and timing-based likelihoods of several astrometric solutions consistent with both techniques to obtain a maximum-posterior astrometric solution. We found offsets between our results and the timing-based astrometric solutions, which, if real, would lead to absorption of spectral power at frequencies of interest for single-source GW searches. However, we do not find significant power absorption due to astrometric fitting at the low-frequency domain of the GW background. 

\end{abstract}



\section{Introduction} \label{sec:intro}

\subsection{Astrometric Parameters in Timing Models}

Pulsar timing allows probing a myriad of astrophysical phenomena by comparing the observed times of arrival (TOAs) of pulses emitted by radio pulsars with predictions from sophisticated mathematical models. These models account for the delays introduced by each of these effects, describing them in terms of a series of parameters. Some of those parameters have been extensively studied over the years, such as the influence of Solar System planetary ephemerides \citep[e.g.,][]{Lazio:2017fos, Vallisneri_2020, Dai_2024}, Solar wind \citep[e.g.,][]{2021A&A...647A..84T, 10.1093/mnras/stae2727}, and relativistic effects \citep[e.g.,][]{2020NatAs...4...72C}, among others. Other parameters are not known a priori or are constrained with limited precision after the discovery of a pulsar, and need to be refined by iteratively including them as free parameters when fitting the timing model to observed TOAs. The latter category includes pulsar astrometric parameters, which comprise:

\begin{itemize}
    \item The sky \textit{position}; in an equatorial coordinate system, this is given by the pulsar's right ascension ($\alpha$) and declination ($\delta$).
    \item The \textit{parallax} angle ($\varpi$).
    \item The \textit{proper motion}; in equatorial coordinates, it is given by the pulsar's proper motion in right ascension ($\mu_{{\alpha}}\equiv\dot{\alpha} \cos{\delta}$) and in declination ($\mu_\delta\equiv\dot{\delta}$).
\end{itemize}

Precise estimations of astrometric parameters are crucial for the precision of pulsar timing fits. In particular, the position and proper motion principally affect the Roemer delay, which accounts for the difference in geometric path length between the pulse arrival time at the observatory and at the Solar System barycenter (SSB). This delay is given by $\Delta_R=-\boldsymbol{r} \cdot \hat{\boldsymbol{R}}_p/c$, where $\boldsymbol{r}$ is the vector pointing from the SSB to the observatory (we neglect errors in
the topocentric to geocentric transformation) and $\hat{\boldsymbol{R}}_p$ is the unit vector from the observatory to the pulsar at the time of the observation \citep[e.g.,][]{2004hpa..book.....L}.

If we ignore any acceleration of the pulsar, approximate the orbit of the Earth as circular, expand to linear order in proper motion and parallax, and drop any constant terms, the (time-variable terms of) the Roemer delay can be written as \citep{2013ApJ...777..104M}

\begin{eqnarray}\label{eq:roemer}
&&\Delta_R\approx\frac{R}{c}\left[\cos{\beta_0}\cos{(\omega t-\lambda_0)}\right.\nonumber -\mu_\beta t\sin{\beta_0}\cos{(\omega t-\lambda_0)}\nonumber\\
&& -\mu_\lambda t\cos{\beta_0}\sin{(\omega t-\lambda_0)} \left.-\tfrac{\varpi}{2}\cos^2{\beta_0}\cos{(2\omega t-2\lambda_0)} \right]
\end{eqnarray}

\noindent where $t$ is a time of conjunction, $\beta(t)=\beta_0+\mu_\beta t$ and $\lambda(t)=\lambda_0+\mu_\lambda t$ are the pulsar's coordinates in an ecliptic system (center at the SSB), $\mu_\lambda$ and $\mu_\beta$ are the components of the proper motion, $\omega$ is $2\pi~\mathrm{yr}^{-1}$, and $R$ is $1$ AU. Real timing analysis utilizes high-precision Solar System ephemerides, rather than a circular approximation of the Earth's orbit, resulting in a more complex dependency on the astrometric parameters.

We see in Eq.~\ref{eq:roemer} that even small offsets in the astrometric parameter estimates from their true values can lead to substantial errors in the Roemer delay calculation. In turn, these errors produce sinusoidal variations in the timing residuals with periods of 1 yr and 6 months (position and parallax, respectively). For example, an error of just $0.1$ arcseconds in the position can cause sinusoidal residual fluctuations with an amplitude of $200~\mu \mathrm{s}$. In turn, these residual excesses can significantly hinder our ability to detect lower-amplitude processes, including gravitational wave (GW) signatures, because they induce delays with the same frequencies for all pulsars, so a parallax or proper motion signal could be misidentified as a GW signal \citep[e.g.,][]{Agazie_2023b}.


Due to the sensitivity of the timing model to such biases, accurate astrometric estimates play an essential role in almost all high-precision pulsar timing experiments. This includes studies of compact binary evolution \citep{Deller_2008}, the distribution of neutron stars in the galaxy \citep{Cordes_1997}, the mapping of interstellar gas \citep{2002astro.ph..7156C}, the properties of globular clusters with pulsars \citep[e.g.,][]{1992RSPTA.341...39P}, and even the neutron star emission mechanisms \citep{Deller_2009}. Pulsar proper motion estimates are also extremely valuable for constraining the origin of pulsars' velocity and es \citep[e.g.,][]{2001ApJ...557L.105F}. 
Moreover, the astrometric distance estimates are vital, together with other independent distance estimates, for calibrating the pulsar distance scale \citep[e.g.,][]{1996ASPC..105..447W, Verbiest_2012}. Of recent and increasing relevance, pulsar distances play a key role in determining the phase of the ``pulsar term'' in GW searches through pulsar timing arrays (PTAs), as this phase is only well-determined if the uncertainty in the pulsar distance is much smaller than the GW wavelength \citep{2023PhRvD.108l3535K}. Usually, PTAs target GWs with a wavelength of about 1 pc--10 pc, so to determine the phase of the pulsar term in the observed signal precisely enough, the distance of the pulsar must be determined with a precision better than about 1 pc.


\subsection{Mis-estimations and Power Absorption}\label{sec:astrometric_biases}

Traditional timing techniques estimate pulsar astrometric values by including them as free parameters when fitting a timing model to observed pulse TOAs. The differences between observed and modeled TOAs are quantified by the residuals. Systematic structures in the residuals with non-random deviations from zero are indicative of parameter mis-estimations. In that case, the timing model is refined by iteratively adjusting its parameters using least-squares fitting. This process continues until the residuals resemble a white noise distribution, resulting in more precise parameter estimates. This fit is usually carried out using software like TEMPO2 \citep{2006MNRAS.372.1549E} or PINT \citep{Luo_2021}.

When performed correctly, this timing-based fit can determine pulsar astrometric parameters with exceptional precision. For normal pulsars, positions can typically be measured with uncertainties of less than an arcsecond. In the case of millisecond pulsars (MSPs), long-term timing campaigns with microsecond TOA precision can achieve sub-milliarcsecond positional accuracy \citep[e.g.,][]{Splaver_2005, 2006MNRAS.369.1502H, Verbiest_2008}. The most precise example to date is PSR J0437$-$4715, for which timing observations have constrained the positional uncertainty to $\lesssim 1~\mathrm{\mu as}$ \citep[e.g.,][]{2024ApJ...971L..18R, Verbiest_2008}.

Nevertheless, pulsar timing-based astrometric fits are susceptible to biases. First, each additional fitted parameter reduces the degrees of freedom in the dataset and introduces potential correlations between parameters. Since models with more free parameters inherently fit data better \citep{Phillips_2022}, including astrometric parameters in the timing fit increases the number of TOAs necessary to maintain statistical robustness and avoid overfitting.

Furthermore, timing-based estimates of pulsar astrometric parameters can be significantly biased when the fit is performed in the presence of red noise, i.e., noise processes with greater power at lower Fourier frequencies than at higher ones. Such processes include spin noise (fluctuations in the pulsar's rotation caused by internal dynamics or changes in magnetospheric torque) which typically follows a power-law spectrum, $P(f) \propto f^{-\gamma}$, with spectral index $4 \lesssim \gamma \lesssim6$ for different pulsars \citep[e.g.,][]{Shannon_2010}. Another major source of red noise is the signature of a stochastic gravitational wave background (GWB) which, if arising from purely GW-driven inspiralling SMBHBs, would manifest as a red-noise-like process in timing residuals with a power spectrum $P(f) \propto f^{-13/3}$ \citep{Jenet_2006, 2008MNRAS.390..192S, 2010arXiv1010.3785C}.


As is the case with other effects, red noise signals introduce characteristic structures in timing residuals. The problem arises when standard least-squares fitting procedures minimize residual amplitudes without distinguishing between the source of such amplitudes. In that case, power from red noise processes can be inadvertently absorbed into astrometric parameter adjustments, obscuring the presence of such processes while also shifting the fitted parameters of the timing model away from their true value \citep[e.g.,][]{10.1111/j.1365-2966.2011.19413.x}. In other words, fitting the timing model by minimizing the root-mean-square (RMS) of the timing residuals causes some power in the red processes to be absorbed into the timing model at the cost of shifting the timing model parameters to possibly erroneous values. 

Since GW signals behave as a red noise process, they too can be partially absorbed into the astrometric model, introducing biases in both astrometric and GW parameter estimates and reducing the overall timing accuracy. Furthermore, astrometric power absorption
results in a reduction of the sensitivity to red noise processes at specific frequencies, some of which coincide with the emission frequency of GW sources, such as the galaxy 3C66B \citep{2020ApJ...900..102A}.

While astrometric signals have characteristic periods of 6 months and 1 year, power from underlying noise processes are not only absorbed at those equivalent frequencies, but power can move between frequencies as a result of the timing model fit (see the timing-model-marginalized residual formalism used in GW searches and the quantification of this movement of power between frequencies $f$ and $f'$ in the matrix $\mathcal{N}^{-1}(f, f')$ in \citealt{hrs2019}). Red noise processes coupled with our astrometric fit can therefore bias not only the astrometric parameters themselves or GW parameter estimates for searches at those frequencies, but at a potentially larger set of frequencies searched over.

To mitigate these biases, alternative fitting algorithms have been developed to account for correlated red noise in pulsar timing data. One such method involves Cholesky whitening techniques \citep[e.g.,][]{2011MNRAS.418..561C}, which reduce the extent to which noise is absorbed into timing model parameters. These techniques improve parameter estimation accuracy and provide more realistic error estimates. However, as proven by \cite{2013ApJ...777..104M}, they do not drastically enhance astrometric parameter precision until enough years of observations have been accumulated such that astrometric errors can manifest distinctly from red noise in the residuals.

\subsection{Synergies and Limitations of VLBI}\label{sec:VLBI_limitations}

Beyond common single-dish timing observations, pulsars have been targeted by Very Long Baseline Interferometry (VLBI) studies since the early 1970s \citep[e.g.,][]{1973ApJ...180L..27V, 1975AJ.....80..923C}.
This technique enables the determination of pulsar positions with sub-milliarcsecond accuracy for certain pulsars, thus achieving precision comparable to the best results obtained from timing-based methods. VLBI astrometry can attain this level of precision after a single observation, whereas timing-based methods typically require around two years of data collection \citep[e.g.,][]{NG15}. Proper motion and parallax can be measured to $10~\mathrm{\mu as}~\mathrm{yr}^{1}$ and $10~\mathrm{\mu as}$, respectively, after approximately 8 epochs over 2 years \citep{Chatterjee_2009}, which is similar to the time baseline required by timing methods.


Notably for our interests, VLBI astrometric measurements are independent of timing data, making them a valuable tool to complement or even enhance timing-based estimates. While timing-based astrometric precision can outperform that from interferometry when a long enough observing baseline is available, there are specific scenarios where using VLBI proves advantageous. This is the case of pulsars close to the ecliptic, particularly at low ecliptic latitudes, whose timing fits are performed in equatorial coordinates; in that case, $\alpha$ and $\delta$ are highly correlated, as are $\mu_\alpha$ and $\mu_\delta$ \citep[see the case of PSR J1022+1001 in][]{2017MNRAS.469..425W}. Depending on their ecliptic latitude, for some pulsars the solar wind electron density is highly covariant with the position and parallax, resulting in significant biases in the timing-derived estimates \citep{10.1093/mnras/stae2727}. Finally, VLBI-informed positions can help refine the kinematic corrections to pulsar period derivatives \cite[see Eq.~4 in][]{2016ApJ...818...92M} and serve as reliable comparison points for the parallax estimates obtained from spindown calculations \cite[see the case of PSR J1024$–$0719 in][]{2016ApJ...818...92M}.

In addition to mitigating correlations, interferometric measurements can assist in establishing an initial timing solution for newly discovered pulsars whose timing data sets span less than two years, which is typically insufficient time to disentangle position, proper motion, and parallax in the timing
analysis. They can also help priors on astrometric parameters that are poorly constrained due to a high rms of the timing residuals.

In particular, VLBI has the potential to improve the determination of astrometric parameters for many pulsars currently included in PTAs. If VLBI and timing observations are obtained over the same time period and combined effectively, the attenuation of red noise amplitude due to model fitting (see Sec.~\ref{sec:astrometric_biases}) could be mitigated. In the case of a red noise process with a spectral index $\gamma$ near $13/3$ for a power law $P(f)\propto f^{-\gamma}$ (as expected from a stochastic GWB from an ensemble of supermassive black hole binaries) and for pulsars with short time baselines ($\sim$500 days of observations), \cite{2013ApJ...777..104M} showed that if we simultaneously fit for the spin and astrometric parameters then post-fit rms is attenuated by a factor of about 20 compared to when we only fit for spin parameters and exclude the astrometric fit. Modern GW searches typically use significantly longer timing datasets ($\sim$5000 days); in this case, the attenuation in rms is the same as in the case where we also fit for the astrometric parameters. Even then, the independence of VLBI estimates from timing measurements could serve as a statistical prior to reduce fit biases caused by parameter covariances.

Beyond enhancing PTA efforts in detecting GWs, VLBI-based measurements of a pulsar's proper motion and parallax could significantly reduce biases in post-Keplerian parameter estimates caused by the Shklovskii effect \citep{1969AZh....46..715S}. This, in turn, would aid in studying the relativistic evolution of pulsar binary systems \citep[e.g.,][]{2009Sci...323.1327D}. Furthermore, VLBI's rapid, high-precision astrometry could improve the localization of newly discovered pulsars, refining their positions following initial radio detections. This improved localization would be instrumental in guiding follow-up observations in X-ray and gamma-ray bands \citep[e.g.,][]{2012MNRAS.422.1294G, 2013ApJ...773L..12B}, thereby enabling multi-wavelength studies of pulsars.

Despite these benefits, different obstacles have hindered previous efforts to integrate VLBI-based astrometric estimates into pulsar timing models:

\begin{itemize}
    \item The timing precision required for pulsar timing experiments can only be achieved with MSPs. Yet, MSPs are relatively faint radio sources, resulting in a limited number of VLBI astrometric measurements for this class of pulsars. Instead, most interferometric studies have focused on young pulsars, which 
    are typically much brighter than MSPs but lack the long-term stability and timing precision required for pulsar timing. The limited number of MSPs that are both bright enough for VLBI studies and stable enough for pulsar timing results in small sample sizes to integrate both techniques.
    \item The astrometric estimates derived by NANOGrav from timing fits 
    are determined in an ecliptic reference frame defined using positions of bodies in the Solar System and tied to the Earth's orbit. In contrast, VLBI measurements
    are determined in an equatorial reference frame defined using distant quasi-static quasars. As a result, VLBI astrometric estimates will suffer a rotational offset from their corresponding timing positions. This offset can sometimes be comparable to the sub-milliarcsecond errors in the estimates, thus rendering VLBI measurements in their original reference frame not suitable for high-precision timing applications.
\end{itemize}

In recent years, new studies have expanded the number of pulsars for which precision VLBI-based astrometry is possible by identifying calibration sources suitable for lower-frequency observations, where pulsars tend to be brighter. Of particular importance is $\mathrm{MSPSR}\pi$ \citep{2023MNRAS.519.4982D}, the first project dedicated to VLBI-based astrometry of MSPs. This project not only included a re-analysis of previously published sources \citep[e.g.,][]{2016ApJ...828....8D}, but also implemented more sophisticated processing techniques to improve astrometric precision further. Similarly, PSR$\pi$ \citep{Deller_2019} also targeted 60 pulsars using the VLBA, significantly increasing the number of pulsars with reliable, model-independent distances beyond 2 kpc. 


\cite{2013ApJ...777..104M} examined how fitting astrometric parameters in the presence of red noise processes, including GWs, can lead to absorption of red noise power into the astrometric fit, weakening their signature in the residuals and introducing significant biases in the timing model. 
Most notably, the authors developed a mathematical formalism to calculate a tie between the VLBI and timing reference frames. This work was later refined by \cite{2017MNRAS.469..425W}, who developed a more robust mathematical framework for the frame tie calculation and included a larger sample of pulsars in its calibration. Their approach enabled a more precise estimation of the frame-tie across multiple ephemerides, which represents a key stepping stone in reconciling VLBI- and timing-based astrometric measurements. These advancements present a new opportunity to integrate VLBI and timing astrometric measurements to further expand the limits of our timing precision and GW sensitivity.

In this paper, we build on this body of work by (1) calibrating an improved frame tie between the reference frames used in VLBI and pulsar timing, and (2) using this calibration to incorporate VLBI-based astrometric estimates as priors in timing models and provide enhanced pulsar astrometric estimates. In Sec.~\ref{sec:corrections}, we present the VLBI and timing-based astrometric measurements used in this study, along with the corrections applied to the data. In Sec.~\ref{sec:frame_tie}, we describe the calibration of the frame tie used to transform VLBI-based estimates into the dynamical Solar System frame of pulsar timing. In Sec.~\ref{sec:methodology}, we outline our methodology for incorporating VLBI priors into a maximum-posterior timing solution. In Sec.~\ref{sec:results}, we present the main results, compare them to existing timing-based estimates, and quantify the impact on power absorption. Finally, in Sec.~\ref{sec:conclusions}, we discuss the implications of this work. Processed data products presented here are publicly available\footnote{See \url{https://github.com/sophiasosafiscella/VLBI_timing} for a living version of the code, and \url{https://zenodo.org/records/17188627} for a frozen version.} as of the date this work is published.

\section{Corrections to Astrometric Estimates}\label{sec:corrections}

In this section, we describe the VLBI- and timing-based astrometric measurements of the MSPs used in this work, as well as the reference frames they are referenced to. We also detail the series of corrections applied to this data to enable a direct comparison between the timing and VLBI estimates.

\subsection{VLBI Observations and Reference Frame Corrections}\label{sec:vlbi_observations}

For our choice of pulsars, we surveyed the existing literature on VLBI pulsar astrometry and selected the sources with published VLBI astrometric measurements that are also included in the 15-year dataset of the North American Nanohertz Observatory for Gravitational Waves \citep[NANOGrav;][]{NG15}. The resulting sample is presented in Table~\ref{table:input_data_table}.

All the VLBI measurements in this work employed the VLBI phase-referencing technique ---a differential VLBI technique designed for observing weak sources--- rather than a traditional absolute VLBI measurement method. This technique requires the use of a \textit{phase calibrator}: a bright, well-known radio source located relatively close to the target source that is used to calibrate the phase of the signal across all antennas. For observations conducted at the L-band (1.4 GHz), such as those performed for most pulsars, the distance should be at a maximum of $4$--$5\degree$ \citep{Chatterjee1999VLBAmemo22}. However, the Earth’s ionosphere  delays signals in a wavelength dependent manner, adding additional path length and contributing phase perturbations. The larger that the target-calibrator separation is, the larger those phase perturbations can be, creating a trade-off between angular distance and brightness.

\begin{deluxetable*}{llc r@{}r@{\fs}l r@{}l r@{}r@{\farcs}l r@{}l ccc}\label{table:input_data_table}
\tablenum{1}
\tablecaption{Pulsar astrometric values used as input data in our work. For each pulsar, the first row corresponds to the VLBI-derived values obtained from the literature, after converting them to the RFC-materialized reference frame; the second row corresponds to timing-derived values from NG15, after converting them to equatorial coordinates and applying proper motion corrections to match the observing epochs used in VLBI. In each VLBI entry, we also present the primary phase calibrator and original catalogue. The $\sigma$ values are the 1$\sigma$ errors in the least significant digit.}
\tablewidth{\linewidth}
\tablehead{
\colhead{PSR} & \colhead{ } & \colhead{\makecell{Epoch\\(MJD)}} &
\multicolumn{3}{c}{\makecell{Right\\Ascension}} & \multicolumn{2}{c}{\makecell{$\sigma_\mathrm{RA}$ \\ (mas)}} &
\multicolumn{3}{c}{Declination} & \multicolumn{2}{c}{\makecell{$\sigma_\mathrm{DEC}$\\(mas)}} &
\colhead{\makecell{Phase\\Calibrator}} & \colhead{\makecell{Original\\Catalog}} & \colhead{Reference}
}
\startdata
\multirow{2}*{J0030+0451} & VLBI & 57849 & 00$^\mathrm{h}$30$^\mathrm{m}$ & 27&42502 & 0&.06 & 04$\degr$51$'$ & 39&7159 & 0&.2 & J0029+0554 & RFC & \cite{2023MNRAS.519.4982D} \\
 & NG15 & 57849 & 00$^\mathrm{h}$30$^\mathrm{m}$ & 27&42504 & 0&.46 & 04$\degr$51$'$ & 39&7141 & 1&.13 & & & \\ \hline
\multirow{2}*{J0437$-$4715} & VLBI & 54100 & 04$^\mathrm{h}$37$^\mathrm{m}$ & 15&883198 & 0&.0567 & $-$47$\degr$15$'$ & 09&032573 & 0&.04966 & J0439$-$4522 & ICRF1 & \cite{Deller_2008} \\
& NG15 & 54100 & 04$^\mathrm{h}$37$^\mathrm{m}$ & 15&883078 & 0&.19 & $-$47$\degr$15$'$ & 09&031878 & 0&.55 & & & \\ \hline
\multirow{2}*{J0610$-$2100} & VLBI & 57757 & 06$^\mathrm{h}$10$^\mathrm{m}$ & 13&570848 & 0&.08 & $-$21$\degr$00$'$ & 27&73907 & 0&.2 & J0610$-$2115 & NVSS & \cite{2023MNRAS.519.4982D} \\
& NG15 & 57757 & 06$^\mathrm{h}$10$^\mathrm{m}$ & 13&600318 & 0&.6 & $-$21$\degr$00$'$ & 27&806726 & 0&.66 & & & \\ \hline
\multirow{2}*{J1012+5307} & VLBI & 57700 & 10$^\mathrm{h}$12$^\mathrm{m}$ & 33&43991 & 0&.04 & \phn53$\degr$07$'$ & 02&1110 & 0&.1 & J0619+0736 & RFC & \cite{2023MNRAS.519.4982D} \\
& NG15 & 57700 & 10$^\mathrm{h}$12$^\mathrm{m}$ & 33&43969 & 0&.26 & \phn53$\degr$07$'$ & 02&1117 & 0&.12 & & & \\ \hline
\multirow{2}*{J1022+1001} & VLBI & 56000 & 10$^\mathrm{h}$22$^\mathrm{m}$ & 43&385081 & 0&.1 & \phn12$\degr$54$'$ & 14&82681 & 1&.0 & J0231+1001 & FIRST & \cite{2016ApJ...828....8D} \\
& NG15 & 56000 & 10$^\mathrm{h}$22$^\mathrm{m}$ & 58&035399 & 225&.55 & \phn10$\degr$01$'$ & 54&272500 & 576&.95 & & & \\ \hline
\multirow{2}*{J1024$-$0719} & VLBI & 57797 & 10$^\mathrm{h}$24$^\mathrm{m}$ & 38&65725 & 0&.06 & $-$07$\degr$19$'$ & 19&8014 & 0&.2 & J1028$-$0844 & RFC & \cite{2023MNRAS.519.4982D} \\
& NG15 & 57797 & 10$^\mathrm{h}$24$^\mathrm{m}$ & 38&65724 & 0&.12 & $-$07$\degr$19$'$ & 19&8030 & 0&.15 & & & \\ \hline
\multirow{2}*{J1640+2224} & VLBI & 57500 & 16$^\mathrm{h}$40$^\mathrm{m}$ & 16&74587 & 0&.07 & \phn22$\degr$24$'$ & 08&7642 & 0&.1 & J1641+2257 & RFC & \cite{2023MNRAS.519.4982D} \\
& NG15 & 57500 & 16$^\mathrm{h}$40$^\mathrm{m}$ & 16&74588 & 0&.04 & \phn22$\degr$24$'$ & 08&7636 & 0&.04 & & & \\ \hline
\multirow{2}*{J1643$-$1224} & VLBI & 57700 & 16$^\mathrm{h}$43$^\mathrm{m}$ & 38&16407 & 0&.1 & $-$12$\degr$24$'$ & 58&6531 & 0&.4 & J1638$-$1415 & RFC &\cite{2023MNRAS.519.4982D} \\
 & NG15 & 57700 & 16$^\mathrm{h}$43$^\mathrm{m}$ & 38&16452 & 0&.44 & $-$12$\degr$24$'$ & 58&6465 & 1&.42 & & & \\ \hline
\multirow{2}*{J1713+0747} & VLBI & 52275 & 17$^\mathrm{h}$13$^\mathrm{m}$ & 49&5306 & 0&.1 & \phn07$\degr$47$'$ & 37&51919 & 2&.0 & J1719+0817 & VCS1 & \cite{Chatterjee_2009} \\
& NG15 & 52275 & 17$^\mathrm{h}$13$^\mathrm{m}$ & 49&5307 & 0&.02 & \phn07$\degr$47$'$ & 37&521752 & 0&.03 & & & \\ \hline
\multirow{2}*{J1730$-$2304} & VLBI & 57821 & 17$^\mathrm{h}$30$^\mathrm{m}$ & 21&67969 & 0&.2 & $-$23$\degr$04$'$ & 31&1749 & 0&.5 & J1726$-$2258 & RFC & \cite{2023MNRAS.519.4982D} \\
& NG15 & 57821 & 17$^\mathrm{h}$30$^\mathrm{m}$ & 21&67923 & 1&.15 & $-$23$\degr$04$'$ & 31&1782 & 1&.86 & & & \\ \hline
\multirow{2}*{J1738+0333} & VLBI & 57829 & 17$^\mathrm{h}$38$^\mathrm{m}$ & 53&97001 & 0&.06 & \phn03$\degr$33$'$ & 10&9124 & 0&.1 & J1740+0311 & RFC & \cite{2023MNRAS.519.4982D} \\
& NG15 & 57829 & 17$^\mathrm{h}$38$^\mathrm{m}$ & 53&96985 & 0&.12 & \phn03$\degr$33$'$ & 10&9205 & 0&.15 & & & \\ \hline
\multirow{2}*{J1853+1303} & VLBI & 57846 & 18$^\mathrm{h}$53$^\mathrm{m}$ & 57&31785 & 0&.06 & \phn13$\degr$03$'$ & 44&0471 & 0&.1 & J1852+1426 & RFC & \cite{2023MNRAS.519.4982D} \\
 & NG15 & 57846 & 18$^\mathrm{h}$53$^\mathrm{m}$ & 57&31804 & 0&.04 & \phn13$\degr$03$'$ & 44&0486 & 0&.09 & & & \\ \hline
\multirow{2}*{J1910+1256} & VLBI & 57847 & 19$^\mathrm{h}$10$^\mathrm{m}$ & 09&70165 & 0&.03 & \phn12$\degr$56$'$ & 25&4316 & 0&.06 & J1911+1611 & RFC & \cite{2023MNRAS.519.4982D} \\
& NG15 & 57847 & 19$^\mathrm{h}$10$^\mathrm{m}$ & 09&70163 & 0&.08 & \phn12$\degr$56$'$ & 25&4327 & 0&.17 & & & \\ \hline
\multirow{2}*{J1918$-$0642} & VLBI & 57768 & 19$^\mathrm{h}$18$^\mathrm{m}$ & 48&02959 & 0&.1 & $-$06$\degr$42$'$ & 34&9335 & 0&.2 & J1912$-$0804 & RFC & \cite{2023MNRAS.519.4982D} \\
& NG15 & 57768 & 19$^\mathrm{h}$18$^\mathrm{m}$ & 48&02967 & 0&.03 & $-$06$\degr$42$'$ & 34&9340 & 0&.2 & & & \\ \hline
B1937+21/& VLBI & 46853 & 19$^\mathrm{h}$39$^\mathrm{m}$ & 38&561185 & 0&.18 & \phn21$\degr$34$'$ & 59&13159 & 2&.4 & B1923+210 & ICRF1 & \cite{1996AJ....112.1690B} \\
J1939+2134 & NG15 & 46853 & 19$^\mathrm{h}$39$^\mathrm{m}$ & 38&561331 & 0&.04 & \phn21$\degr$34$'$ & 59&13134 & 0&.09 & & & \\ \hline
\multirow{2}*{J2010$-$1323} & VLBI & 56000 & 20$^\mathrm{h}$10$^\mathrm{m}$ & 45&9211 & 0&.1 & $-$13$\degr$23$'$ & 56&083 & 4&.0 & J2011$-$1546 & RFC & \cite{Deller_2019} \\
& NG15 & 56000 & 20$^\mathrm{h}$10$^\mathrm{m}$ & 45&9212 & 0&.07 & $-$13$\degr$23$'$ & 56&082 & 0&.5 & & & \\ \hline
\multirow{2}*{J2145$-$0750} & VLBI & 56000 & 21$^\mathrm{h}$45$^\mathrm{m}$ & 50&4588 & 0&.1 & $-$07$\degr$50$'$ & 18&513 & 2&.0 & J2142$-$0437 & RFC & \cite{Deller_2019} \\
 & NG15 & 56000 & 21$^\mathrm{h}$45$^\mathrm{m}$ & 50&4588 & 0&.32 & $-$07$\degr$50$'$ & 18&514 & 1&.22 & & & \\ \hline
\multirow{2}*{J2317+1439} & VLBI & 56000 & 23$^\mathrm{h}$17$^\mathrm{m}$ & 9&2364 & 0&.1 & \phn14$\degr$39$'$ & 31&265 & 1&.0 & J2318+2404 & RFC & \cite{Deller_2019} \\
& NG15 & 56000 & 23$^\mathrm{h}$17$^\mathrm{m}$ & 9&2364 & 0&.03 & \phn14$\degr$39$'$ & 31&264 & 0&.16 & & & \\
\enddata
\end{deluxetable*}

Both the astrometric positions of the pulsar and those of the phase calibrator are reported with high precision in the International Celestial Reference System (ICRS). The ICRS is a \textit{reference system}, i.e., the totality of procedures, models, and constants set by the International Earth Rotation and Reference Systems Service (IERS) that specify the orientation of the system's axes. As such, a reference system provides a theoretical definition, which is later implemented using a catalog of the adopted coordinates of a set of reference objects. Such a realization of the theoretical reference system is known as a \textit{reference frame}.

Various astrometric campaigns have realized the ICRS to different levels of precision using catalogs of reference sources observed with VLBI. The current standard, adopted by the International Astronomical Union, is the International Celestial Reference Frame (ICRF), a catalog of equatorial coordinates of 608 extragalactic radio sources. These positions define the ICRS axes to a precision of $\sim 20~\mu\mathrm{as}$ \citep{1998AJ....116..516M}. The ICRF has since had two successors: the ICRF2 includes 3414 sources \citep{Fey_2015}, and the ICRF3 expands this to 4536 sources  \citep{2020A&A...644A.159C}. 
However, the ICRF is not the only realization of the ICRS. The Radio Fundamental Catalogue (RFC), for instance, comprises the absolute positions for 21942 compact radio sources, with typical median position uncertainties ranging from $\sim 0.3$ to $1.0$~mas for well-observed sources, and reaching $0.1$~mas for the most accurately determined cases \citep{Petrov_2025}.

The VLBI campaigns referenced in this work employed phase calibrators drawn from different catalogs, and thus referenced to different realizations of the ICRS (see eighth column of Table~\ref{table:input_data_table}). Before comparing astrometric estimates, it is essential to ensure that all VLBI measurements are placed within a common reference frame. In particular, most of the interferometric positions in our sample were obtained from the MSPSR$\pi$ project, which references its positions to the RFC. As of 2024, the RFC was considered ``the most complete and most precise catalog'' of its kind \citep{Petrov_2025}. For consistency with MSPSR$\pi$, we therefore refer all interferometric positions to the frame realized by the RFC.

In principle, the transformation between different VLBI catalogs corresponds to a three-dimensional rotation between their respective realizations of the ICRS. However, as mentioned before, phase-referenced VLBI pulsar positions are determined relative to nearby calibrator sources. In this context, if the calibrator’s absolute position is updated (because the reference catalog changed), we can propagate that correction to the pulsar just by adding the same offset. Therefore, the frame difference can be represented locally as the difference in the calibrator’s catalogued coordinates, which can then be applied as a linear offset to the pulsar position \citep[as done in][]{2017MNRAS.469..425W}. This procedure is equivalent to the first-order approximation of the global rotation when restricted to a small region of the sky; since the offsets are extremely small (microarcseconds to milliarcseconds), the higher-order spherical terms are negligible (see Appendix \ref{sec:linearized} for a detailed demonstration).

Let $(\alpha^\mathrm{cat},\delta^\mathrm{cat})$ be a pulsar's interferometric position in its original catalog. If this position was calibrated using a source whose equatorial coordinates in that original catalog are $(\alpha^\mathrm{cat}_\mathrm{calib},\delta^\mathrm{cat}_\mathrm{calib})$, and that the same calibration source has coordinates $(\alpha^\mathrm{RFC}_\mathrm{calib},\delta^\mathrm{RFC}_\mathrm{calib})$ in the RFC, then the pulsar's coordinates can be calibrated to the RFC-defined reference frame as:

\begin{equation}\label{eq:catalog_correction}
    \begin{split}
        \alpha^\mathrm{RFC} &= \alpha^\mathrm{cat} + (\alpha^\mathrm{RFC}_\mathrm{calib} - \alpha^\mathrm{cat}_\mathrm{calib})\\
        \delta^\mathrm{RFC} &= \delta^\mathrm{cat} + (\delta^\mathrm{RFC}_\mathrm{calib} - \delta^\mathrm{cat}_\mathrm{calib})
    \end{split}
\end{equation}

In most cases, the known positional uncertainties in the calibration sources used are included in the reported $\alpha$ and $\delta$ uncertainties. In those cases, we did not correct the published uncertainties to those in the RFC because we cannot separate the calibration error from other errors. However, for PSR J0437$-$4715, both uncertainties were given independently, so we summed in quadrature: (a) the uncertainty in the calibrator source, (b) the original positional uncertainty, and (c) a calibration error of $\sim 0.8$~mas in the \cite{Deller_2008}'s right ascension of PSR J0437$-$4715 (reported in \citealt{2017MNRAS.469..425W}).

\subsection{Timing Observations and Epoch Corrections}\label{sec:timing_observations}

For the timing-based astrometric estimates, we used the timing solutions reported in the NANOGrav 15-year data set release (hereafter NG15), as presented in Table~\ref{table:input_data_table}. These solutions were derived from observations conducted at the Arecibo Observatory, the Green Bank Telescope, and the Very Large Array, covering frequencies from 327 MHz to 3 GHz, with most sources observed approximately once per month. The timing analysis and subsequent model fitting were performed using PINT \citep{Luo_2021} as the primary timing software. All fits adopted the JPL DE440 Solar System ephemeris \citep{Park_2021} and the TT(BIPM2019) timescale. Full details on data collection, calibration, pulse arrival-time determination, and timing model creation for NG15 are provided in \cite{NG15}.

The reference system used in pulsar timing is inherently tied to the orientation of Earth's orbit relative to the coordinate axes of a specified Solar System planetary ephemeris. Consequently, the resulting astrometric measurements depend on the choice of ephemeris. For example, \citet{1984MNRAS.210..113F} showed that timing positions derived using the MIT ephemeris differ systematically by approximately $0.2$~arcseconds from those based on a JPL ephemeris. Ensuring consistency in the ephemeris is therefore essential before comparing timing-based positions. In principle, timing positions can be transformed between ephemerides via rotation matrices \citep[e.g.,][]{1996AJ....112.1690B}. However, since all timing solutions in NG15 used the JPL DE440 ephemeris, no additional corrections were necessary.

The astrometric solutions published in NG15 are provided in ecliptic coordinates; for consistency with the VLBI astrometric solutions, we used \texttt{PINT}'s \texttt{as\_ICRS} method to transform them to equatorial coordinates. Moreover, the NG15 timing positions are referenced to specific observing epochs that do not necessarily coincide with the epochs of the corresponding interferometric measurements. These mismatches can result in substantial positional differences due to proper motions. For instance, PSR J0437$-$4715, with a proper motion of $\mu \sim 140.915(1)~\mathrm{mas/yr}$ \citep{Verbiest_2008}, would exhibit a positional shift of $\sim 1\farcs53598(1)$ over the $\sim 10.9$ year interval between the VLBI and timing observations. Therefore, it is essential to account for proper motion during these epoch offsets before comparing VLBI and timing astrometric measurements. To that end, we used the \texttt{change\_posepoch} method from the \texttt{PINT} timing software to update the timing astrometric positions of each pulsar to the corresponding epochs used in the VLBI observations through the following calculation:

\begin{equation}
    \begin{split}
    \alpha^\mathrm{DE}_\mathrm{(new)} = \alpha^\mathrm{DE}_\mathrm{(old)} + \mu_\alpha \Delta t + \varpi f_\alpha (\alpha^\mathrm{DE}_\mathrm{(old)}, \delta^\mathrm{DE}_\mathrm{(old)}, \Delta t) \\
    \delta^\mathrm{DE}_\mathrm{(new)} = \delta^\mathrm{DE}_\mathrm{(old)} + \mu_\delta \Delta t + \varpi f_\delta (\alpha^\mathrm{DE}_\mathrm{(old)}, \delta^\mathrm{DE}_\mathrm{(old)}, \Delta t)
    \end{split}
\end{equation}

\noindent where ($\alpha^\mathrm{DE}_\mathrm{(old)}, \delta^\mathrm{DE}_\mathrm{(old)}$) are the pulsar's right ascension and declination in the original epoch, ($\alpha^\mathrm{DE}_\mathrm{(new)}, \delta^\mathrm{DE}_\mathrm{(new)}$) are the same parameters in the new epoch, $\mu_\alpha$ and $\mu_\delta$ are the proper motions in the two respective directions, $\Delta t$ is the time difference between the two epochs, $\varpi$ is the parallax, and $f_\alpha$ and $f_\delta$ are the parallax displacements of the pulsar in right ascension and declination, which can be expressed as:

\begin{align}\label{eq:proper_motion_correction}
    f_{\alpha} = \sin \lambda \cos \alpha \cos \varepsilon& - \cos \lambda \sin \alpha \nonumber \\
    f_{\delta} = \sin \lambda \sin \varepsilon \cos \delta 
    &- \sin \lambda \cos \varepsilon \sin \alpha \sin \delta  \\
    &- \cos \lambda \cos \alpha \sin \delta \nonumber
\end{align}

\noindent where $\lambda$ is the solar elliptical longitude and $\varepsilon$ is the obliquity of the ecliptic. For a full discussion on these corrections and ensuing correlations, see \cite{universe11020054}.

\section{Reference System Considerations}\label{sec:frame_tie}

As explained in Sec.~\ref{sec:VLBI_limitations}, a major challenge in integrating VLBI and timing astrometric measurements is the difference in the reference systems used by each technique. In particular:

\begin{itemize}
    \item \textbf{VLBI observations} are referenced to the ICRS, a barycentric, non-rotating reference system whose axes are fixed with respect to distant radio sources that have negligible proper motions \citep{2005jsrs.meet...19A}. As such, the ICRS is independent of Earth's dynamics and is considered a \textit{kinematic} system, satisfying the kinematic condition for an inertial system in general relativity. This theoretical definition is materialized through catalogs of reference sources observed in radio using VLBI, such as the ICRF.
    \item \textbf{Pulsar timing observations} are referred to a \textit{dynamic} reference system, also centered on the Solar System barycenter, but oriented relative to the Earth's orbital plane. As such, this system is strictly tied to the orientation of Earth's orbit with respect to the coordinate axes of a given Solar System planetary ephemeris, such as the JPL's latest ephemeris, the DE440.
\end{itemize}

The relative orientation of these two systems depends on the position and inclination of the Earth's spin axis. As a result, astrometric measurements of the same object, but referenced to either system, will suffer from a systematic offset due to the rotation between the axes of the corresponding systems. This effect has been well-studied by several authors. Work as early as \cite{1984MNRAS.210..113F} has reported systematic discrepancies of $0.2$--$0.5$ arcseconds between VLBI-based positions and those derived from timing using JPL's DE96 ephemeris. Similarly, \cite{1996AJ....112.1690B} found a systematic offset of $\sim10$~mas when using JPL's DE200. More recently, \cite{2017MNRAS.469..425W} compared the positional measurements of PSR J0437$-$4715 across different ICRS realizations and planetary ephemerides, and found an offset of $\sim1$~mas even for the modern ICRF2 and DE435. Finally, \cite{Park_2021} reports that the orientations of the inner planets reported in the DE440 ephemeris are aligned with the ICRF3 with an average accuracy of about 0.2 mas.


An early attempt to characterize this offset was made by \citet{1994A&A...287..279F}, who performed an indirect determination of the tie between the extragalactic (VLBI-based) and planetary (timing-based) reference frames. Their realization of the ICRS was based on a 14-year-long VLBI catalog from NASA's Deep Space Network, while the timing frame was constructed using ephemerides of the inner planets, constrained by lunar laser ranging data. However, modern realizations of both systems have improved substantially. The current ICRS realization, the ICRF3, incorporates a larger number of sources observed over several decades of VLBI data. Similarly, current timing efforts rely on more comprehensive planetary ephemerides, such as the JPL DE440, which also account for the effects of the outer planets and other refinements. An updated determination of the VLBI–timing frame tie is therefore both timely and necessary.

An updated frame tie estimation was calculated by \cite{universe11020054}. In this work, the authors used the published astrometric parameters of 23 MSPs with VLBI observations referenced to the ICRF3, and timing observations from the International Pulsar Timing Array (IPTA) data release two \citep[DR2;][]{10.1093/mnras/stz2857}. As a result, they obtained estimations of the frame ties between the ICRF3 and two dynamical Solar System frames ---one using the DE436 and another using the DE440 ephemerides--- with an accuracy exceeding $0.4$ mas. In contrast, our work uses a different sample of MSPs, with timing observations from the NG15 data set (instead of the IPTA DR2) and VLBI observations referenced to the RFC (instead of the ICRF3). We will thus perform a completely independent frame tie calculation using a different calibration data set, which can then be compared to \cite{universe11020054}'s results.

\subsection{Mathematical Formalism}\label{sec:frame_tie_math}

\begin{figure}[ht]
\centering
    \includegraphics[width=0.6\linewidth]{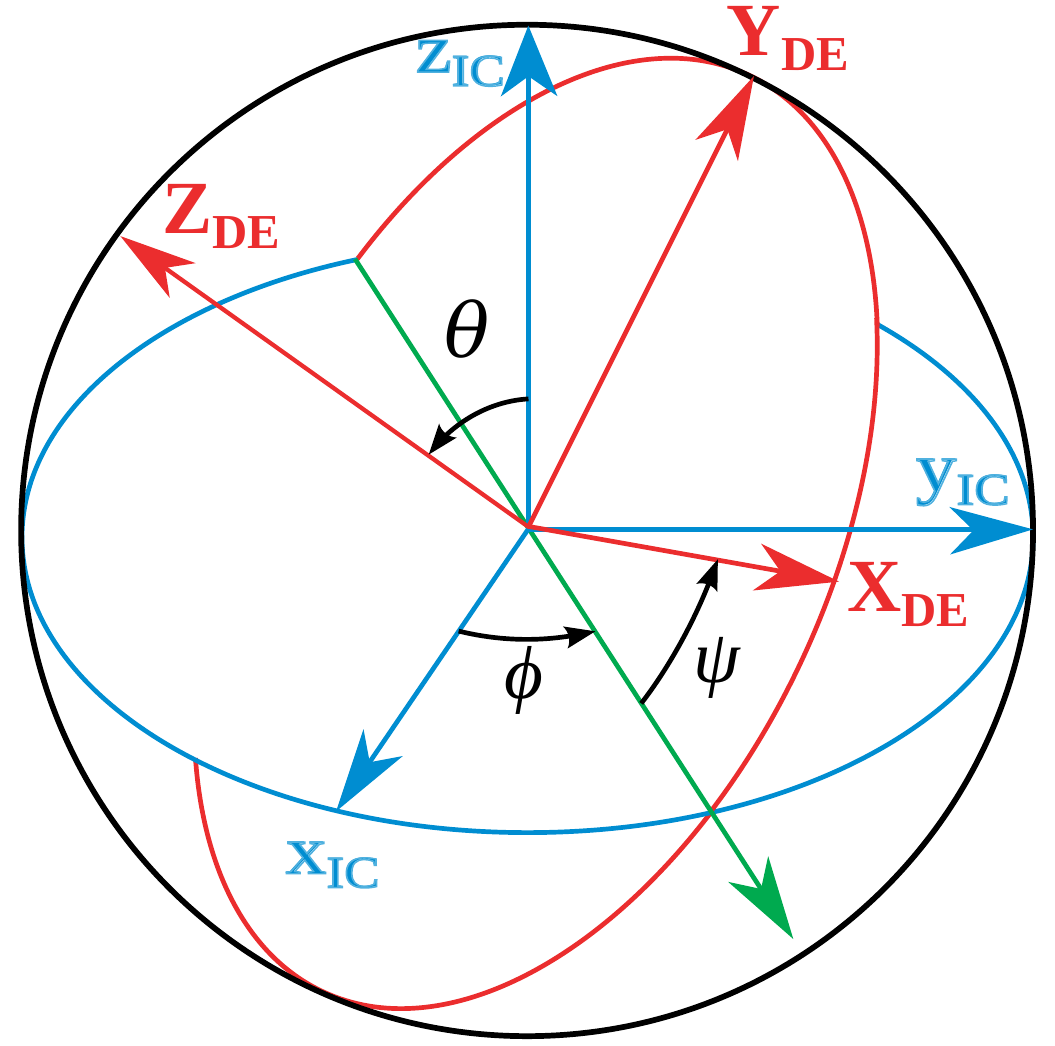}
\caption{Celestial sphere diagram showing the sequence of Euler rotations, $\boldsymbol{R}_z(\phi) \boldsymbol{R}_x (\theta) \boldsymbol{R}_z(\psi)$, that transform the ICRS (IC) into the timing reference system (DE). The magnitude of the Euler angles $\phi,\theta,\psi$ is exaggerated for visualization purposes; in practice, they are smaller than $10^8$ rad.}\label{fig:sphere}
\end{figure}

Since the ICRS and the timing reference system share the same origin and differ only by the orientation of their axes, their relationship can be expressed as three consecutive rotations. These can be represented through the matrix product $\boldsymbol{\Omega} = \boldsymbol{R}_z(\phi) \boldsymbol{R}_x (\theta) \boldsymbol{R}_z(\psi)$, where $\boldsymbol{R}_k$ is the elementary rotation matrix around the $k$-axis and $\phi,\theta,\psi$ are the Euler rotation angles (see Fig.~\ref{fig:sphere}). $\boldsymbol{\Omega}$ is thus a $3 \times 3$ orthogonal rotation matrix whose entries are trigonometric polynomials in $\phi,\theta,\psi$. If a pulsar has coordinates $\widehat{\textbf{n}}^\mathrm{IC}$ in the ICRS and $\widehat{\textbf{n}}^\mathrm{DE}$ in the timing system, then

\begin{equation}\label{eq:frame_tie_1}
    \widehat{\textbf{n}}^\mathrm{DE}_i + \boldsymbol{\epsilon}^\mathrm{DE}_i= \boldsymbol{\Omega} (\widehat{\textbf{n}}^\mathrm{IC}_i + \boldsymbol{\epsilon}^\mathrm{IC}_i)
\end{equation}

\noindent where the index $i$ indicates the $i^\mathrm{th}$ pulsar and the $\boldsymbol{\epsilon}$ terms represent measurement errors.

Since the rotation angles and the measurement errors, $\boldsymbol{\epsilon}_i$, are small ($< 10^{-8}$ rad), we can use the small-angle approximation to linearize the trigonometric polynomials in $\boldsymbol{\Omega}$. In that case, the rotation matrix becomes $\boldsymbol{\Omega} \approx \textbf{I} +\textbf{A}$, where \textbf{I} is the identity matrix and $\textbf{A}$ is an antisymmetric matrix given by

\begin{equation}
    \textbf{A} = 
    \begin{pmatrix}
    0 & A_z & -A_y\\
    -A_z & 0 & A_x\\
    A_y & -A_x & 0
    \end{pmatrix}
\end{equation}

\noindent where $A_x, A_y, A_z\ll1$ are infinitesimal rotations around the fixed Cartesian axes $x$, $y$, $z$. Using this approximation, \cite{2013ApJ...777..104M} derived a linear least squares problem, 

\begin{equation}\label{eq:frame_tie}
    \widehat{\textbf{n}}^\mathrm{DE}_i - \widehat{\textbf{n}}^\mathrm{IC}_i = \textbf{m}_i \widetilde{\textbf{A}}+\boldsymbol{\Sigma}
\end{equation}

\noindent Here, $\widetilde{\textbf{A}}^T=[A_x~A_y~A_z]$ is the rotation vector that produces the frame tie, $\boldsymbol{\Sigma}$ is the covariance matrix of $\boldsymbol{\epsilon}^\mathrm{IC}_i - \boldsymbol{\epsilon}^\mathrm{DE}_i$, and

\begin{equation}
    \textbf{m}_i =
    \begin{pmatrix}
        0 & -\widehat{n}^\mathrm{IC}_{i,z} & \widehat{n}^\mathrm{IC}_{i,y} \\
        \widehat{n}^\mathrm{IC}_{i,z} & 0 & -\widehat{n}^\mathrm{IC}_{i,x} \\
        -\widehat{n}^\mathrm{IC}_{i,y} & \widehat{n}^\mathrm{IC}_{i,x} & 0 \\
    \end{pmatrix}
\end{equation}

\cite{2013ApJ...777..104M} applied this mathematical formalism to determine the rotation angles between the VLBI and timing reference frames using two pulsars: PSR J0437$-$4715 (the closest and brightest MSP) 
and PSR J1713+0747 (one of the pulsars most extensively studied for timing). Using their reported values\footnote{The values reported in \cite{2013ApJ...777..104M} are $(A_x, A_y, A_z) \approx(2.26$~mas, $-0.36$~mas, $2.26$~mas); we confirmed with the authors that these values might be affected by biases.} of $A_x, A_y, A_z$ to transform the interferometric positions in Table~\ref{table:input_data_table} into the timing reference frame, we found offsets as large as $2$~arcseconds when compared to NG15 timing positions at the same reference epochs. Moreover, in our independent implementation of their algorithm, using the same set of position measurements for the same two pulsars, we obtained $(A_x, A_y, A_z) \approx(3.34$~mas, $2.70$~mas, $-3.03$~mas). These value do not reproduce their results, and did not provide an improved the agreement between the timing and interferometric positions.

In solving Eq.~\ref{eq:frame_tie}, \cite{2013ApJ...777..104M} assumed that $\boldsymbol{\Sigma}$ is diagonal because VLBI and timing measurement uncertainties are uncorrelated and because they assumed that correlations in position uncertainties for individual pulsars can be ignored. However, \cite{2017MNRAS.469..425W} pointed out that this assumption has two problems: (a) $\boldsymbol{\Sigma}$ is singular and cannot be inverted as required by the solution to Eq.~\ref{eq:frame_tie}, and (b) $\boldsymbol{\Sigma}$ is not diagonal because the measurements in right ascension and declination for individual pulsars have correlations.

In view of these shortcomings, \cite{2017MNRAS.469..425W} avoided the singularity in the covariance matrix $\boldsymbol{\Sigma}$ by considering that the position vector $\widehat{\textbf{n}}=(x,y,z)$ can be related to the equatorial coordinates $\boldsymbol{\theta}=(\alpha,\delta)$ through 

\begin{equation}
    x = \cos{\alpha} \cos{\delta}; \qquad y=\sin{\alpha}\cos{\delta}; \qquad z=\sin{\delta}
\end{equation}

\noindent The differentials can be written as $d\widehat{\textbf{n}}= \textbf{D}_\mathrm{eq} d\boldsymbol{\theta}$, where

\begin{equation}\label{eq:Wang_frame_tie}
    \textbf{D}_\mathrm{eq} =
    \begin{pmatrix}
        -\sin{\alpha} \cos{\delta} & -\cos{\alpha} \sin{\delta} \\
        \cos{\alpha} \cos{\delta} & -\sin{\alpha} \cos{\delta} \\
        0 & \cos{\delta}
    \end{pmatrix}
\end{equation}

Using Eq.~\ref{eq:Wang_frame_tie}, the authors obtained the new least squares problem given by:

\begin{equation}\label{eq:frame_tie_6}
        d\boldsymbol{\theta}^\mathrm{DE} - d\boldsymbol{\theta}^\mathrm{IC} = (\textbf{D}^T_\mathrm{eq} \textbf{D}_\mathrm{eq})^{-1} \textbf{D}^T_\mathrm{eq} \textbf{M} \widetilde{\textbf{A}}+\boldsymbol{\Sigma}_\theta.
\end{equation}

\noindent The covariance matrix here, $\boldsymbol{\Sigma}_\theta$ is the sum of the covariance matrix of the VLBI and timing positions, i.e., $\boldsymbol{\Sigma}_\mathrm{IC} + \boldsymbol{\Sigma}_\mathrm{DE}$. $\boldsymbol{\Sigma}_\mathrm{DE}$ can be extracted from the timing fit performed using packages such as TEMPO2 or PINT. However, the correlations between the $\alpha$ and $\delta$ measurements using VLBI are usually not included in the published results, so $\boldsymbol{\Sigma}_\mathrm{IC}$ needs to be approximated as diagonal. As \cite{2017MNRAS.469..425W} noted, the need for including the off-diagonal terms increases when $\alpha$ and $\delta$ are highly correlated, such as in the case of pulsars very close to the ecliptic.

Eq.~\ref{eq:frame_tie_6} represents a new least squares problem whose results are numerically identical to those from Eq.~\ref{eq:frame_tie}, but they are derived directly from the $(\alpha, \delta)$ measurements rather than the transformed direction cosines $(x,y,z)$. \cite{2017MNRAS.469..425W} calibrated this frame tie using 5 MSPs with position determinations from VLBI measurements referenced to the ICRF2 and timing measurements referenced to 6 different ephemerides. In this work, we will adopt the same mathematical formalism, but we will expand the data set of MSPs used to calibrate the rotation parameters.

\subsection{Frame Tie Calculation}

In addition to using 5 MSPs, \cite{2017MNRAS.469..425W} quantified the effects of also including young (canonical) pulsars in the frame tie calibration. They concluded that, while young pulsars would not systematically bias the resulting frame tie estimation, their contribution is negligible unless the timing uncertainties can be greatly improved. Based on this result, we restrict our frame tie calibration to astrometric measurements of MSPs only, as they not only contribute most significantly to the calibration, but are also the primary targets of pulsar timing array experiments.

For the frame tie calculation, we used 16 of the MSPs listed in Table~\ref{table:input_data_table}. First, we corrected the interferometric positions to account for differences in the reference frame realizations of the source catalogs (see Eq.~\ref{eq:catalog_correction}). For the corresponding NG15 timing position, we applied proper motion corrections to match the VLBI reference epochs (see Eq.~\ref{eq:proper_motion_correction}). We then developed an independent Python implementation of the mathematical framework detailed in \citet{2017MNRAS.469..425W}, and calibrated it using this corrected data. The complete numerical implementation and corresponding calibration data set can be found in our online repositories (see Appendix \ref{sec:appendix}).

Two sources, PSR~J0610$-$2100 and PSR~J1022$+$1001, were excluded from the calibration, since their relatively large positional uncertainties and discrepancies between the VLBI and timing estimates make them not suitable for use in frame tie calculations. In the case of PSR J1022+1001, these uncertainties can be attributed to its low ecliptic latitude, since the position of such pulsars is difficult to fit well into the equatorial coordinates \citep[e.g.:][]{Yan_Shen_Yuan_Wang_Rottmann_Alef_2012, 2017MNRAS.469..425W}. Coincidentally, \cite{universe11020054} also excluded these two pulsars from their frame tie calculation based on similar arguments.

The final calibrated Euler rotation angles defining the frame tie are:

\begin{align}\label{eq:my_frame_tie}
    A_x &= -0.68 \pm 0.16~\mathrm{mas} \nonumber \\
    A_y &= 0.15 \pm 0.61 ~\mathrm{mas} \\
    A_z &= 0.70 \pm 0.63 ~\mathrm{mas} \nonumber
\end{align}

\noindent Given our choice of catalog of VLBI catalog and planetary ephemeris, these angles describe the rotation between the ICRS as realized by the RFC catalog and the dynamical planetary reference frame defined by the JPL DE440 Solar System ephemeris. Frame tie calculations using a different combination of catalog and ephemeris would require transforming all astrometric positions to the corresponding reference frames prior to calibration.

Notably, our frame tie estimates are in full agreement, within uncertainties, with \cite{universe11020054}'s independent calculations of the frame tie between the ICRF3 and the DE440 planetary ephemerides frame, given by $A_x = -0.86 \pm0.21~\mathrm{mas}, A_y=0.13\pm0.17~\mathrm{mas}, A_z=0.00\pm0.25~\mathrm{mas}$. Even though their VLBI astrometric parameters are referred to the ICRF3 and not to the RFC, like in our work, the fact that these results were independently obtained and still agree at the level of precision indicates the remarkable agreement between both realizations of the ICRS, and further corroborates \cite{universe11020054}'s results.

\subsection{Applying the Frame Tie}

\begin{figure*}[ht]
\centering
    \includegraphics[width=0.65\linewidth]{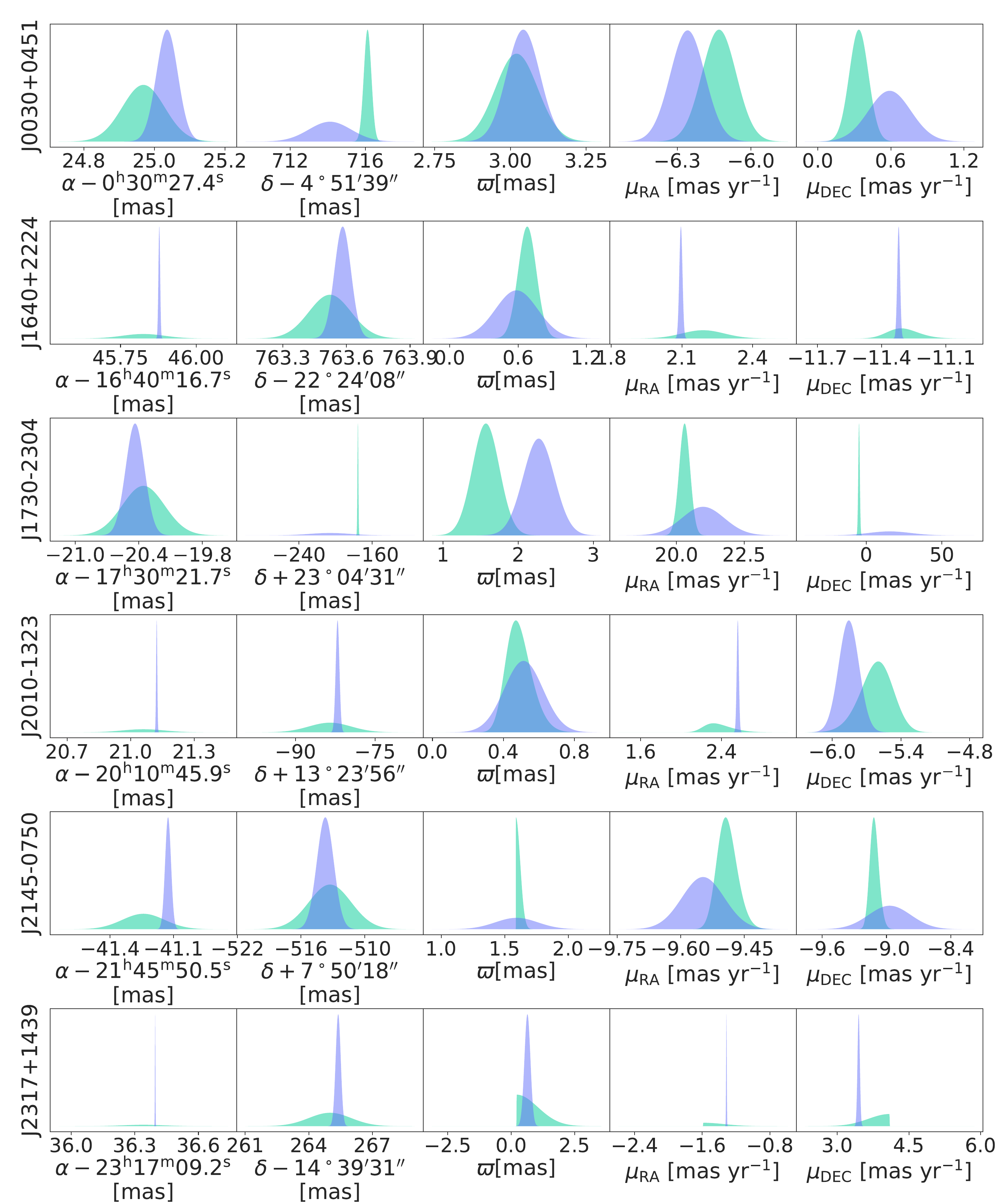}
\centering
   
\caption{PDFs of the astrometric parameters derived using VLBI (green) or timing (blue) for a selected subsample of our pulsars when referred to the same reference frame. The VLBI-derived PDFs are based on the measurements from Table~\ref{table:input_data_table}, now converted to the dynamical Solar System reference frame by means of the frame tie in Eq.~\ref{eq:my_frame_tie}.}\label{fig:PDFs}
\end{figure*}

We used Eq.~\ref{eq:frame_tie_1} to transform the interferometric positions and corresponding errors from Table~\ref{table:input_data_table} into the dynamical reference frame. Fig.~\ref{fig:PDFs} shows the resulting probability density functions (PDFs) for each pulsar, derived from VLBI and timing measurements. When the reported errors were symmetric, the PDF was modeled as a Gaussian distribution. For asymmetric error intervals $[\bar{x} - u_L, \bar{x} + u_R]$, we modeled the PDF with a skew-normal distribution, defined as 

\begin{equation}
    f(x)=\frac{2}{\omega\sqrt{2 \pi}} e^{-\frac{x-\xi}{\omega}}\Phi(\alpha x),
\end{equation}

\noindent where $\Phi(x)$ is the cumulative distribution function and $\alpha$ is the skewness parameter. To estimate the $\xi, \omega, \alpha$ values that best reproduce a distribution with percintiles given by  $\bar{x}-u_L$ and $\bar{x}+u_R$, we performed an independent implementation of the algorithm presented in \cite{Possolo_2019}, which estimates maximum likelihood values of $\xi, \omega, \alpha$ using the Nelder-Mead optimization method \citep{10.1093/comjnl/7.4.308}.

Before applying a frame tie, we found significant discrepancies between the estimates of the position and proper motion components derived using VLBI and timing. However, after applying frame tie \ref{eq:my_frame_tie} to rotate the VLBI astrometric estimates to the dynamical solar
system frame, we now observe a significant overlap between the VLBI- and timing-derived PDFs. These results are presented for a selected subsample of 6 test pulsars in Fig.~\ref{fig:PDFs}. This improved agreement highlights the effectiveness of the frame tie calibration in mitigating the component of the systematic offsets arising from differences in reference frames and enhancing the consistency between VLBI and timing astrometric measurements.

Notably, the VLBI-derived PDFs for PSRs J2317$+$1439, J2145$-$0750, and J2010$-$1323 appear sharply asymmetric. The original distributions obtained by \cite{Deller_2019} were determined using a bootstrap fitting technique: they resampled the time series of measured positions, performed a least-squares fit for each resampled dataset, and extracted the most probable value and associated confidence intervals from the resulting cumulative distribution functions. The resulting uncertainty asymmetry depends on both the number of position measurements and the presence of outliers relative to the best-fit model. For example, a limited number of high-quality observations of PSR J2317$+$1439 caused the bootstrap solutions to cluster asymmetrically in parameter space (A. T. Deller, private communication, 2025), resulting in highly skewed, conservative uncertainty estimates. Additionally, \cite{Deller_2019} reported astrometric values based on the mode of the bootstrap distribution \citep[see Fig. 4 in][]{Deller_2019} rather than the median. In cases where the distributions are strongly non-Gaussian, this choice can further exaggerate the reported asymmetry. Subsequent work from the MSPSR$\pi$ project \citep{2023MNRAS.519.4982D} adopted a Bayesian inference framework instead, yielding more robust and symmetric uncertainties, but it did not include updated measurements for these three pulsars.

\section{Methodology}\label{sec:methodology}

In this section, we describe our procedure for generating trial timing solutions by sampling different combinations of the astrometric parameters for each pulsar in our dataset. We also outline how VLBI-based astrometric measurements (corrected for catalog differences and transformed to the timing reference frame) were incorporated as priors in the construction of the posterior distributions for these timing solutions.

\subsection{Samples Selection}\label{sec:timing_samples}

\begin{figure}[ht]
\centering
    \subfigure[Sampling in parallax.]{\includegraphics[width=0.9\linewidth]{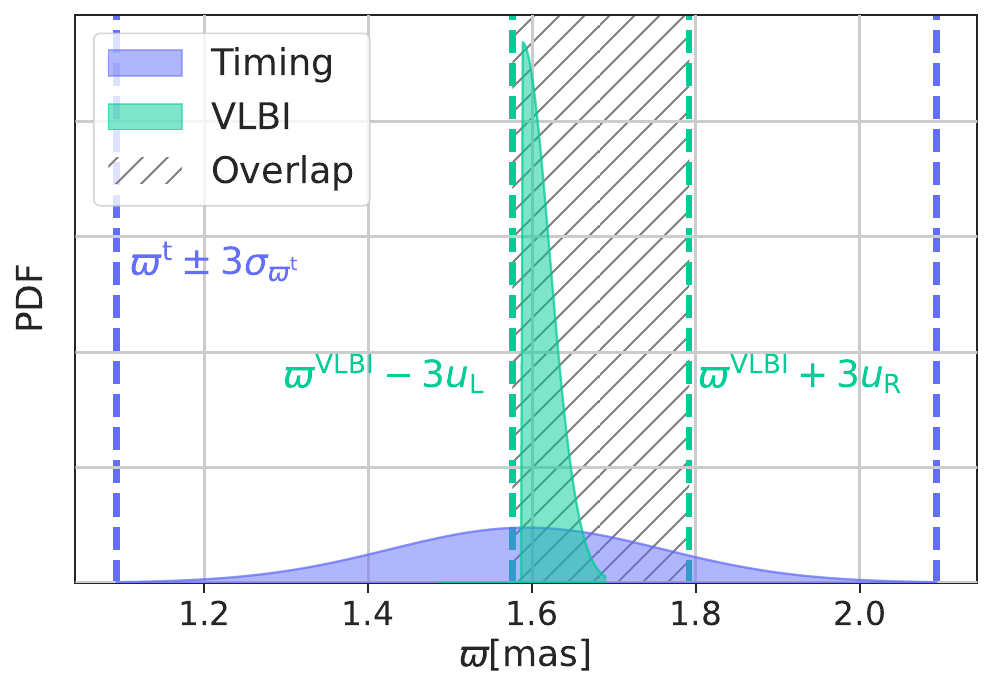}\label{subfig:parallax_sampling}}
\centering
\subfigure[Sampling in proper motion.]{\includegraphics[width=\linewidth]{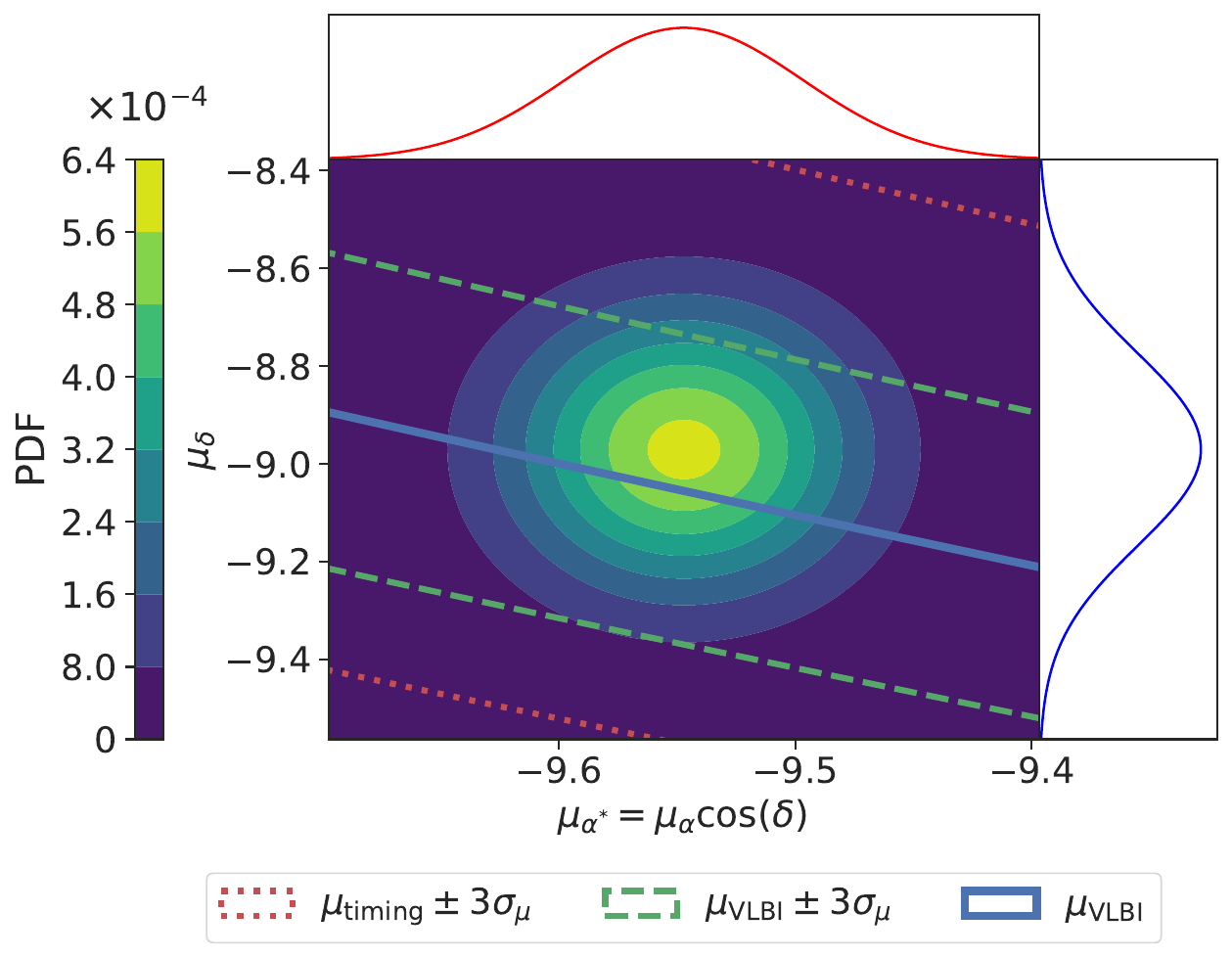}\label{subfig:proper_motion_sampling}}
   
\caption{Parameter sampling for PSR J2145$-$0750. Panel~\ref{subfig:parallax_sampling} shows the PDFs derived from the VLBI- (green) and timing- (blue) based parallax estimations. The vertical dashed lines correspond to a $3\sigma$ deviation from the mean value of each distribution. The sampled $\varpi_i$ values are selected from the overlap of these ranges. In panel~\ref{subfig:proper_motion_sampling}, the side plots show the marginal distributions in $\mu_\alpha$ and $\mu_\delta$ derifved from timing, and the central color plot is the 2D joint distribution of both parameters. The solid green line represents the total proper motion derived from VLBI, and the dashed green lines on each side are its $3\sigma$ deviations; the sampled $({\mu_{\alpha}}_i , {\mu_{\delta}}_i)$ pairs are selected from this range.
\label{fig:sampling}}
\end{figure}

After transforming the VLBI measurements from the RFC reference frame to the dynamical planetary frame, our goal is to use these VLBI-derived astrometric estimates as priors to compute a maximum-posterior timing solution. To do this, we first generated a set of trial timing solutions by sampling various combinations of the astrometric parameter values. Since the parameter space is continuous and infinite, exhaustively exploring all possible values is not computationally feasible. Instead, for each pulsar and each astrometric parameter, we sampled values that were consistent with both the timing and the VLBI constraints as follows:

\begin{itemize}
    \item For the right ascension,  we calculated the overlap of the VLBI ($\mathrm{V}$) and timing ($\mathrm{t}$) PDFs within a $3\sigma$ range of the mean value of the distribution, i.e.,

    \begin{equation}\label{eq:3sigma_ranges}
    [\bar{\alpha}^{(\mathrm{t})} \pm 3\sigma_{\alpha^{(\mathrm{t})}}] \cap [\bar{\alpha}^{(\mathrm{V})} \pm 3\sigma_{\alpha^{(\mathrm{V})}}]
    \end{equation}

    \noindent We then selected $N$ equidistant values $\{\alpha_i\}_{i=1,\ldots,N}$ in this overlap. Given the finite resolution, we adopted the difference between two consecutive samples as a measurement of the statistical uncertainty.

    \item For the declination, we proceeded analogously to the right ascension.

    \item For the parallax, if the error bar estimates were symmetric, we proceeded analogously to the right ascension. However, when the error bars were asymmetric, we sampled $\{\varpi_i\}_{i=1,\ldots,N}$ values in 
    
    \begin{equation*}
    [\bar{\varpi}^\mathrm{(\mathrm{t})} \pm 3\sigma_{\varpi^\mathrm{t}}] \cap [\bar{\varpi}^\mathrm{(V)} - 3 u^L_{\varpi}, \bar{\varpi}^\mathrm{(V)}+3 u^R_{\varpi}]
    \end{equation*}
    
    \noindent where $u^L_{\varpi}$ and $u^R_{\varpi}$ and the left and right error bars of the VLBI-derived parallax estimates (see Fig.~\ref{subfig:parallax_sampling}).
    
   \item Since the total proper motion is a vector, its magnitude, $\mu$, is independent of the reference system. Therefore, to mitigate biases introduced by the frame conversion, we compared interferometric and timing values of the total proper motion:

    \begin{enumerate}
        \item We computed the VLBI-based total proper motion, $\mu^\mathrm{(V)} = \left( {\mu^\mathrm{(V)}_\alpha}^2+{\mu^\mathrm{(V)}_\delta}^2 \right)^{1/2}$, and its error bars, $u_{\mu^\mathrm{(V)}}^{L}$ and $u_{\mu^\mathrm{(V)}}^{R}$, through propagation of the uncertainties in $\mu^\mathrm{(V)}_\alpha$ and $\mu^\mathrm{(V)}_\delta$.
        \item Then, we selected $\{{\mu_{\alpha}}_i\}_{i=1,\ldots,N}$ samples of $\mu^\mathrm{(t)}_\alpha$ and $\{{\mu_{\delta}}_j\}_{j=1,\ldots,N}$ samples of $\mu^\mathrm{(t)}_\delta$ in their corresponding $3\sigma$ ranges, analogously to Eq.~\ref{eq:3sigma_ranges}.
        \item We kept the $\left({\mu_{\alpha}}_i^\mathrm{(t)},{\mu_{\delta}}_j^\mathrm{(t)}\right)$ pairs such that

        \begin{equation}\label{eq:proper_motion_sampling}
        \begin{split}
        \left( {{\mu_{\alpha}}_i^\mathrm{(t)}}^2 + {{\mu_{\delta}}_j^\mathrm{(t)}}^2 \right)^{1/2} &\in \\ [\bar{\mu}^\mathrm{(V)} - 3 u^L_{\mu^{(V)}} &, \bar{\mu}^\mathrm{(V)} + 3 u^R_{\mu^{(V)}}]
        \end{split}
        \end{equation}
        
        \noindent This process is illustrated in Fig.~\ref{subfig:proper_motion_sampling}.
        
    \end{enumerate}

\end{itemize}

Finally, we obtained trial astrometric solutions $(\alpha_j, \delta_j, \varpi_j, {\mu_{\alpha}}_j, {\mu_\delta}_j)$ by taking all possible combinations of the sampled values for each parameter, i.e., $\{\alpha_i\} \times \{\delta_i\} \times \{\varpi_i\} \times  \{({\mu_{\alpha}}_i , {\mu_{\delta}}_j)\} $. The number of total possible combinations can be as large as $N^5$ if all $N^2$ pair of $({\mu_{\alpha}}, {\mu_\delta})$ satisfy Eq.~\ref{eq:proper_motion_sampling}, but it can be as low as $N^3$ if only one pair satisfies it. Therefore, we sampled at most $N=8$ values per parameter, giving a largest-case scenario of $32 768$ possible different solutions. 

\subsection{Timing Likelihood Calculation}

We then calculated the likelihood that the pulsar TOAs reported in NG15 are described by the timing model corresponding to each trial astrometric solution found in Sec.~\ref{sec:timing_samples}:

\begin{enumerate}
    \item \textbf{Astrometric substitution}: Starting from the NG15 timing solution, we replaced its astrometric parameter values with those corresponding to the astrometric solution $(\alpha_j, \delta_j, \varpi_j, {\mu_{\alpha}}_j, {\mu_\delta}_j)$ in question.
    \item \textbf{Initial timing fit}: Using the \texttt{PINT} software, we performed an initial fit of the modified model to the NG15 TOAs. The 5 astrometric parameters were held fixed, while all other timing parameters were allowed to vary.
    \item \textbf{Noise model optimization}: We re-fit the noise model using a Parallel Tempering MCMC sampler from the \texttt{PINT pal} repository\footnote{\url{https://github.com/nanograv/pint\_pal}}. This algorithm samples different combinations of the parameters that describe white noise processes (\texttt{WN}) and red noise (\texttt{RN}). 
    We then selected the maximum likelihood noise realization using Bayesian inference and inserted it into the timing model.
    \item \textbf{Final fit}: We performed a second fit of the timing model with the updated noise parameters to NG15's TOAs. Again, we fixed the astrometric parameters and fitted all other timing parameters. The likelihood $L$ of the trial astrometric solution was then calculated by examining the post-fit residuals. In the presence of correlated noise, this likelihood is given by \citep[][]{Susobhanan_2024}:

    \begin{equation}\label{eq:likelihood}
        \ln{L}=-\frac{1}{2} \textbf{r}^T \textbf{C}^{-1} \textbf{r} -\frac{1}{2} \ln{\det{\textbf{C}}}
    \end{equation}

    \noindent where $\textbf{r}$ are the post-fit timing residuals and $\textbf{C}$ is the non-diagonal covariance matrix that incorporates both white and correlated noise.


\end{enumerate}

\begin{figure}[ht]
\centering
    \includegraphics[width=\linewidth]{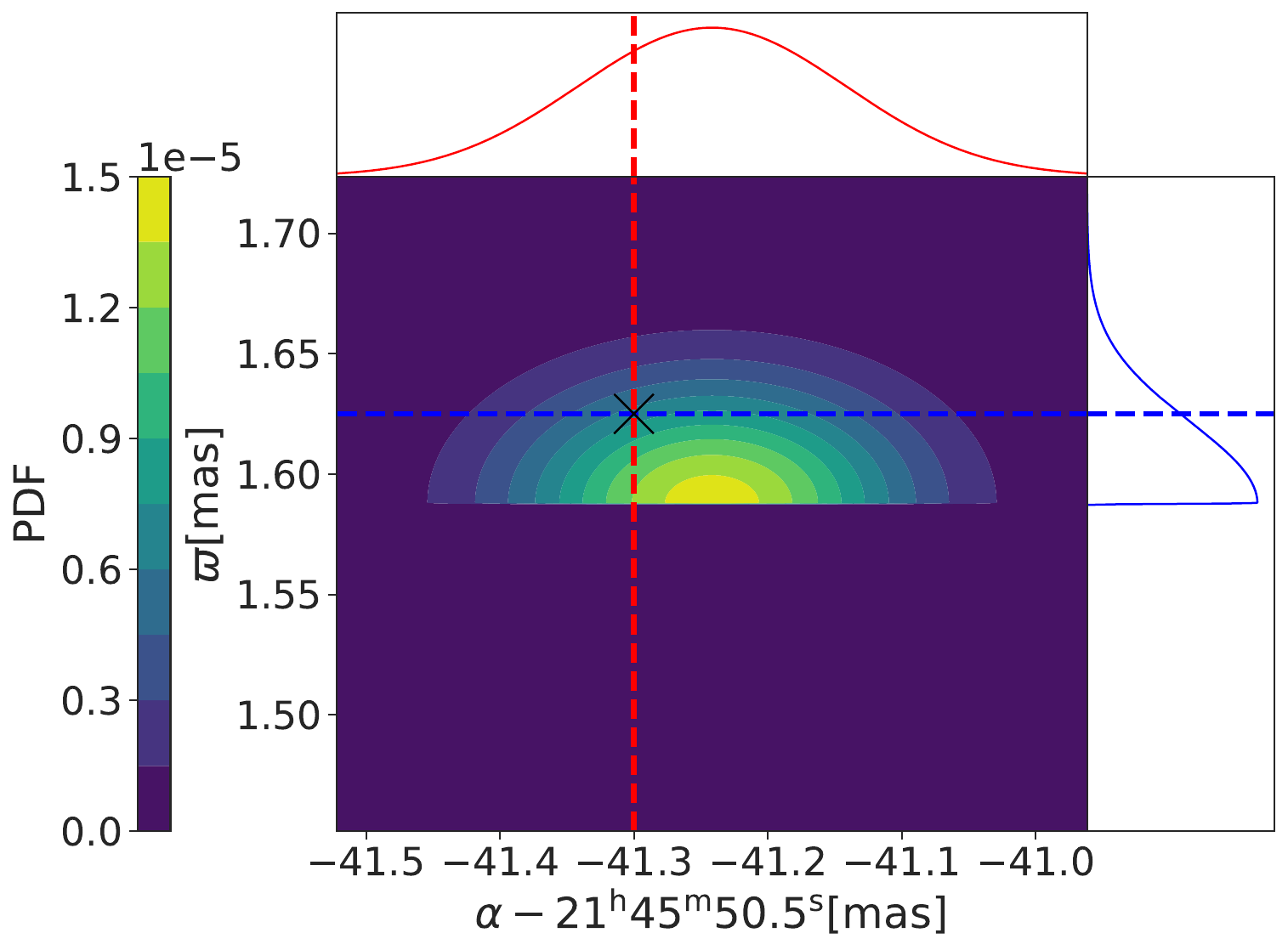}
\caption{2-dimensional example of prior calculation for a trial timing solution of J2145$-$0750 with $(\alpha,\varpi)=(21^\mathrm{h}45^\mathrm{m}50\fs4586,1.625~\mathrm{mas})$. The side plots show the VLBI-derived PDF of the parameter on the corresponding axis. When evaluated on the trial parameter value (dotted lines), this PDF gives the corresponding parameter prior. The total prior is given by the joint probability of these values (black cross on color map).}\label{fig:priors}
\end{figure}

\subsection{VLBI Prior Calculation}
For each pulsar, we used its VLBI-based astrometric estimates to assign a prior probability to each trial astrometric solution. In particular, for each sampled value of the astrometric parameters, we evaluated the value of its VLBI-derived PDF at the sampled point. For example, if $f^\mathrm{(V)}_\alpha(x)$ is the VLBI-derived PDF for the right ascension, then the prior for a sampled value $\alpha_j$ is simply $f^\mathrm{(V)}_\alpha(\alpha_j)$. The total prior for the full astrometric solution $s_j\equiv(\alpha_j, \delta_j, \varpi_j, {\mu_{\alpha}}_j, {\mu_\delta}_j)$ was then computed as the joint probability of the individual priors:

\begin{equation}\label{eq:prior}
    f(s_j) = f^\mathrm{(V)}_\alpha(\alpha_j) f^\mathrm{(V)}_\delta(\delta_j) f^\mathrm{(V)}_\varpi(\varpi_j) f^\mathrm{(V)}_{\mu_\alpha}({\mu_\alpha}_j) f^\mathrm{(V)}_{\mu_\delta}({\mu_\delta}_j)
\end{equation}

\noindent This process is illustrated in 2 dimensions in Fig.~\ref{fig:priors}.

\subsection{Posteriors Calculation}

Once the likelihood and prior have been derived for each trial timing solution $s_j$, we calculate the corresponding posterior using Bayes' Theorem as:

\begin{equation}
    \mathrm{Pr}(s_j|D) = \frac{\mathrm{Pr}(D|s_j) \mathrm{Pr}(s_j)}{\mathrm{Pr}(D)}
\end{equation}

\noindent where $D$ refers to the pulsar's NG15 TOA dataset, $\mathrm{Pr}(s_j|D)$ is the posterior probability of the astrometric solution $s_j$, $\mathrm{Pr}(D,s_j)\equiv L(s_j)$ is the likelihood (given by Eq.~\ref{eq:likelihood}), and $\mathrm{Pr}(s_j)$ is the prior probability (given by Eq.~\ref{eq:prior}). The Bayesian evidence, $\mathrm{Pr}(D)$, acts as a normalization factor that is independent of the parameters; therefore, we can ignore it for a maximum likelihood analysis.

\begin{figure*}[ht]
\centering

\subfigure[J0030$+$0451]{\includegraphics[width=0.48\textwidth]{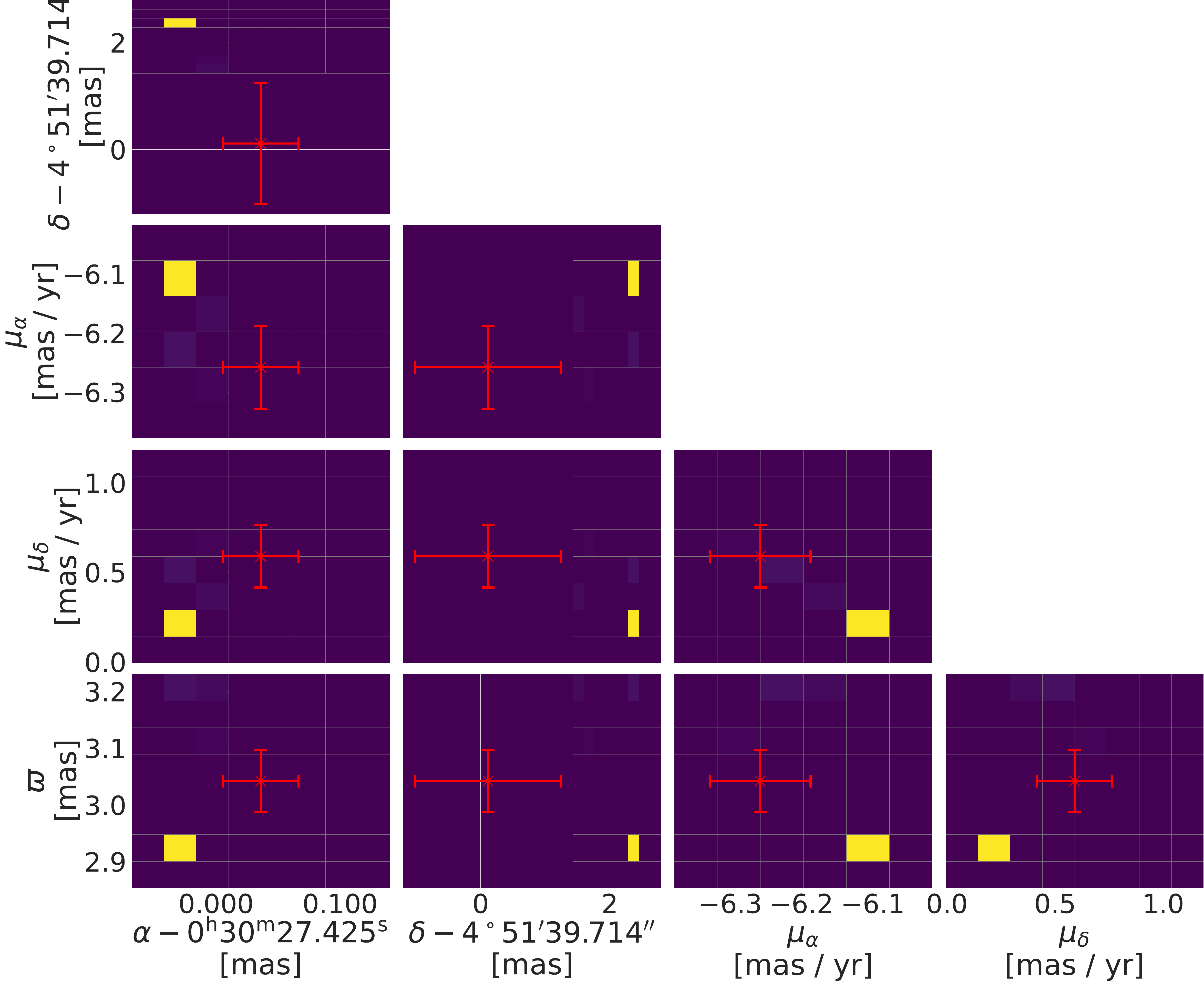}\label{fig:J0030_cornerplot}}
\subfigure[J1640$+$2224]{\includegraphics[width=0.48\textwidth]{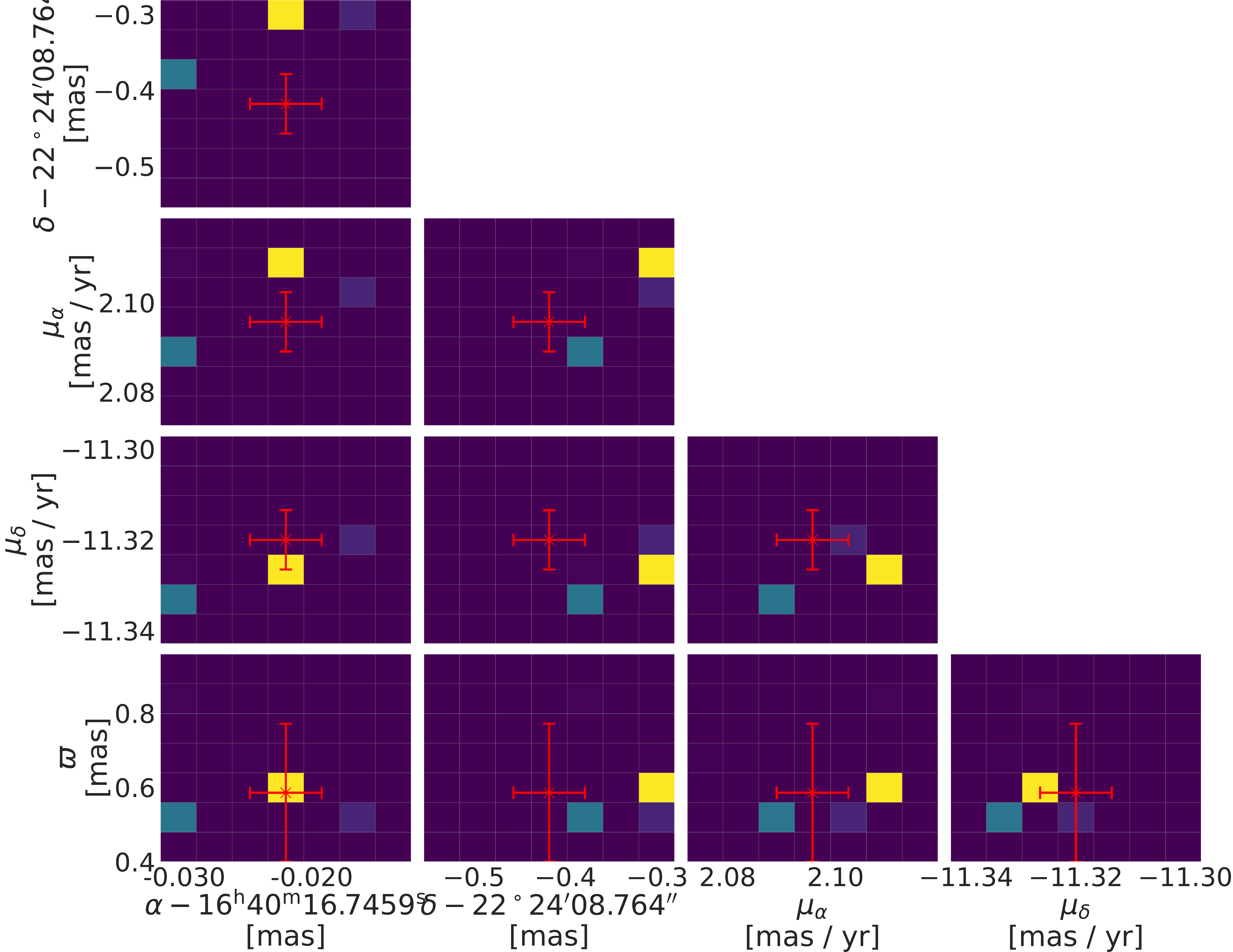}\label{fig:J1640_cornerplot}} \\[-2ex]

\subfigure[J1918$-$0642]{\includegraphics[width=0.48\textwidth]{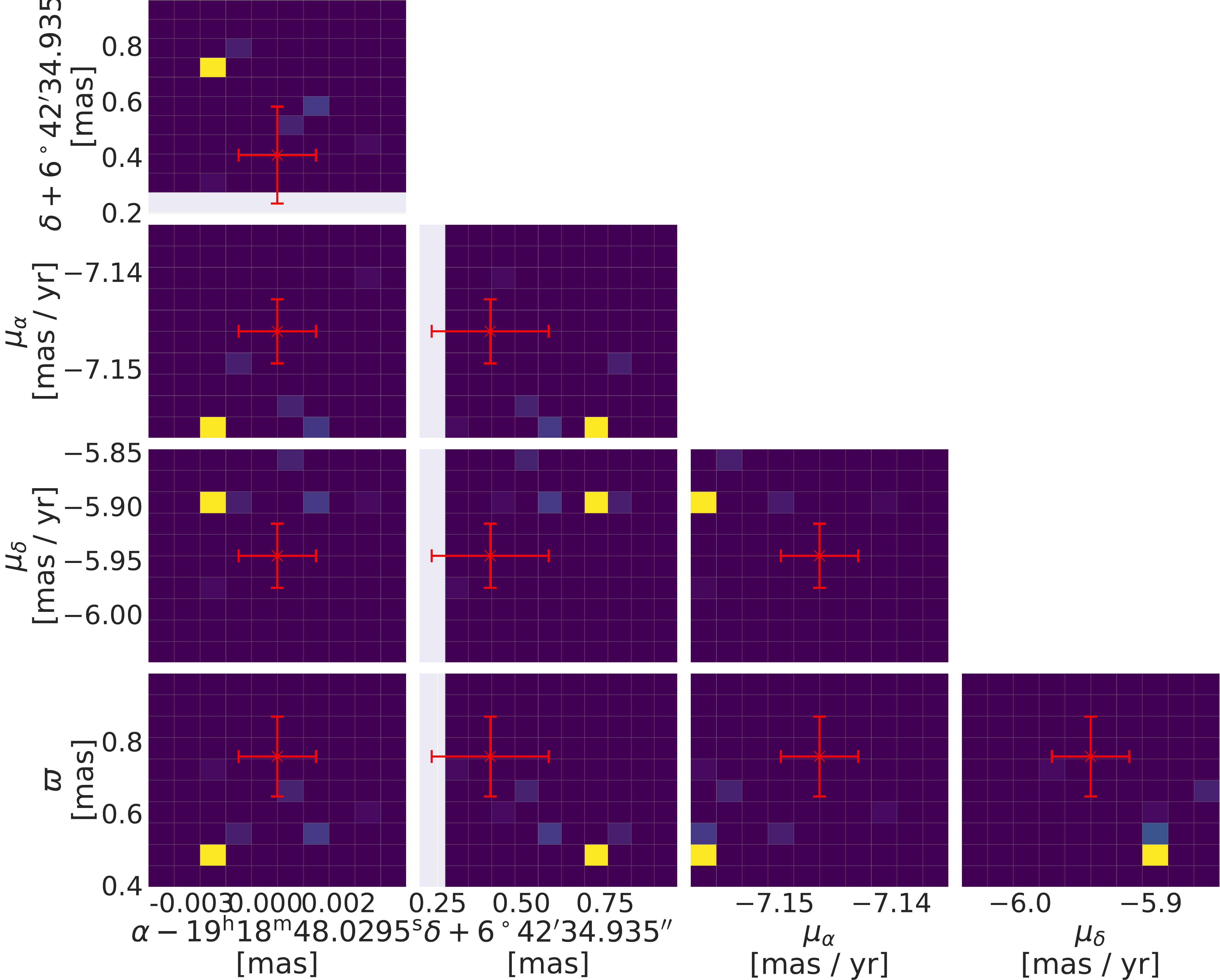}\label{fig:J1730_cornerplot}}
\subfigure[J2010$-$1323]{\includegraphics[width=0.48\textwidth]{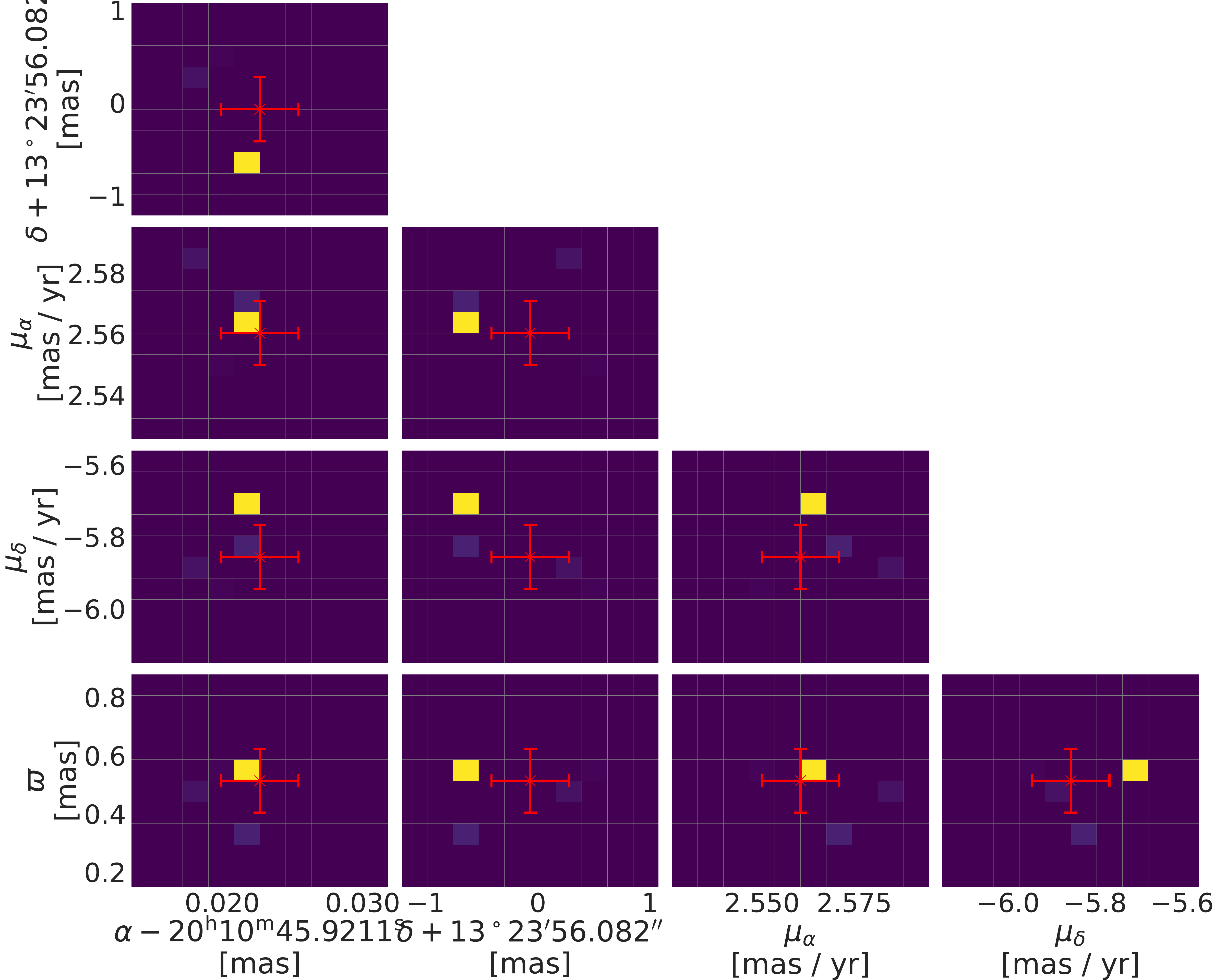}\label{fig:J2010_cornerplot}} \\[-2ex]

\subfigure[J2145$-$0750]{\includegraphics[width=0.48\textwidth]{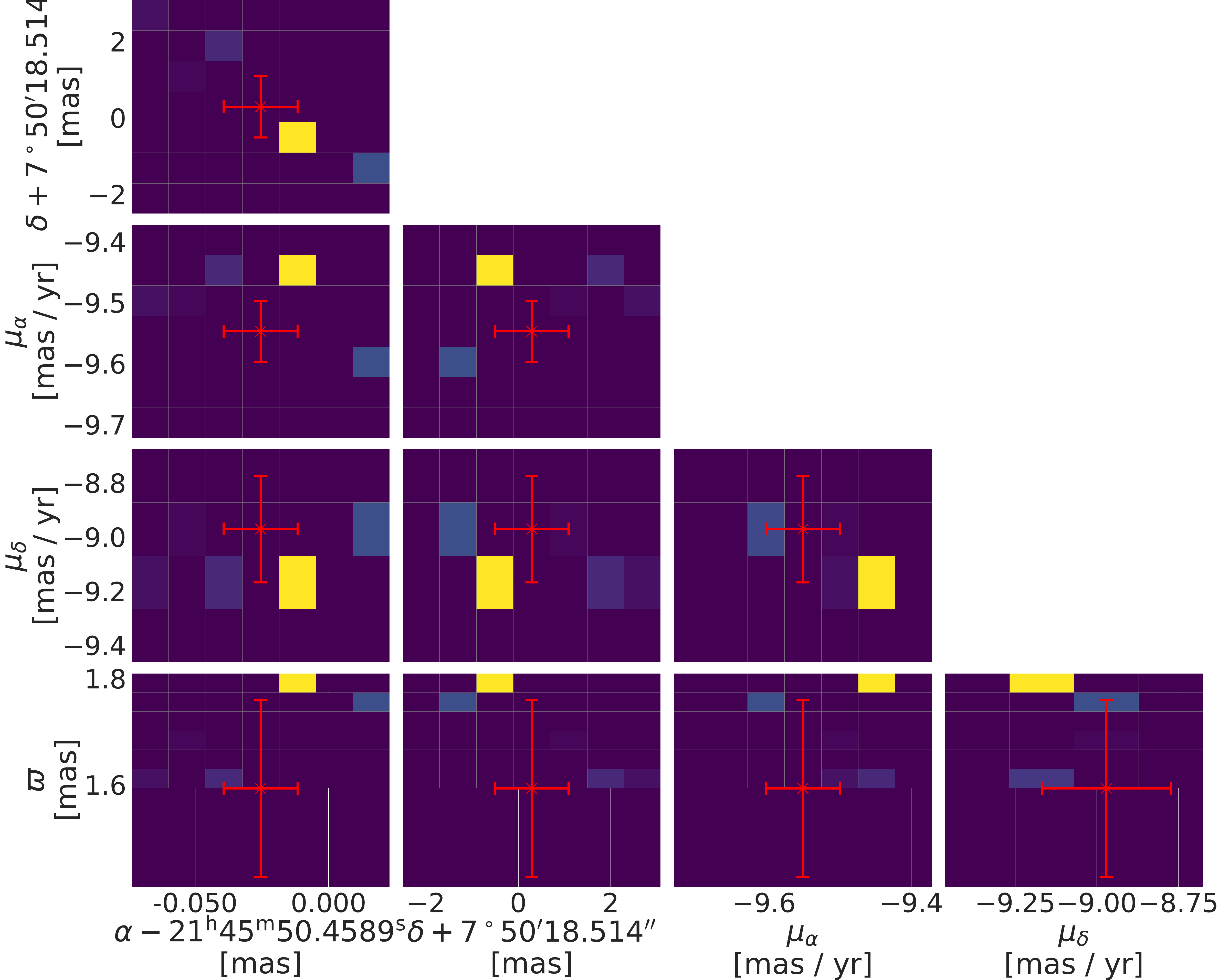}\label{fig:J2145_cornerplot}}
\subfigure[J2317$+$1439]{\includegraphics[width=0.48\textwidth]{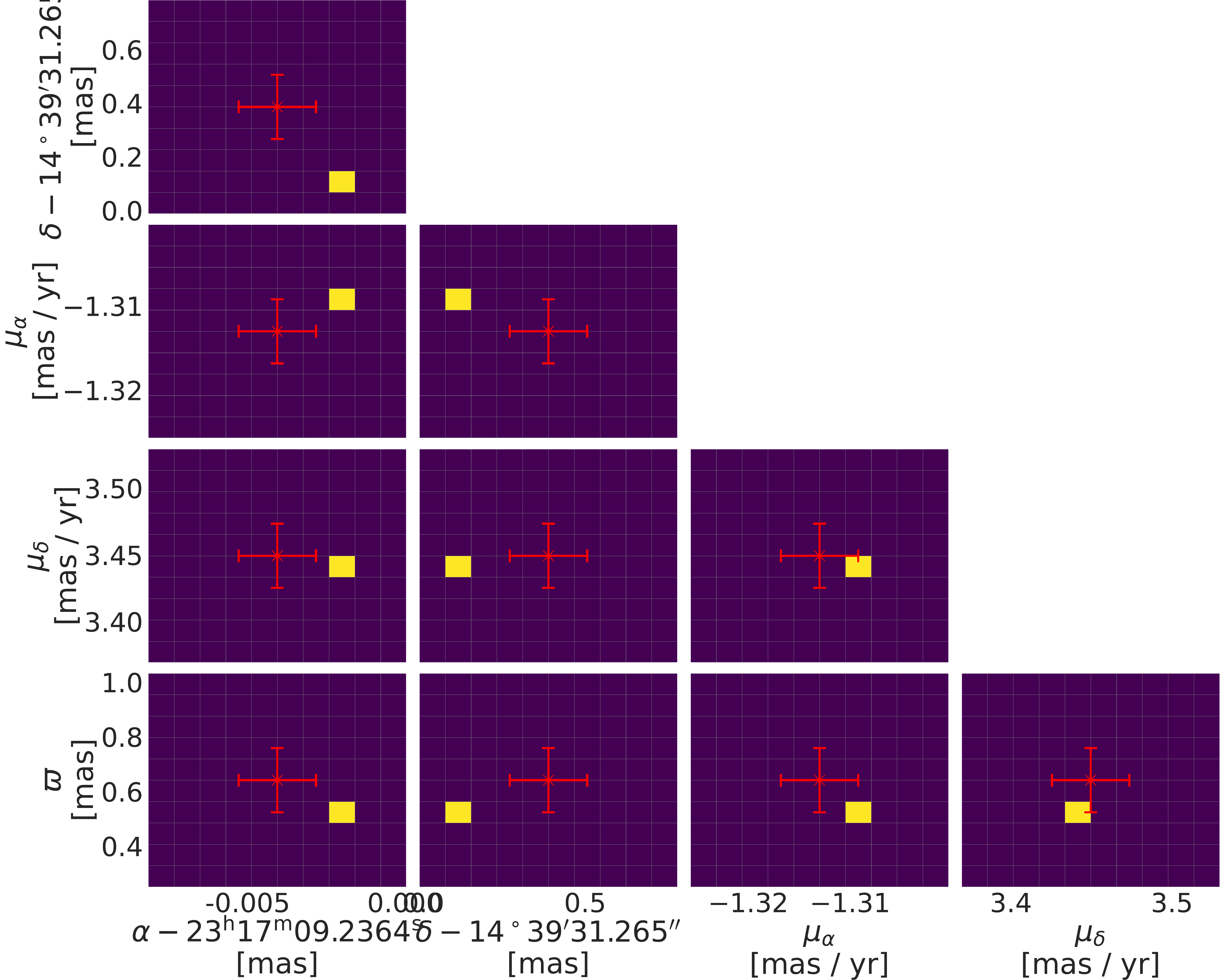}\label{fig:J2317_cornerplot}}

\caption{Corner plots of posterior probability. In each subplot, the colormap displays the normalized posterior distribution (yellow: higher posterior, purple: lower posterior) as a function of one pair of astrometric parameters, after marginalizing the posteriors over the three other astrometric parameters. Grey areas correspond to regions of the parameter space that were not sampled in this work. 
Each bin is a trial astrometric solution. The red error bars show the NG15's timing astrometric solution. 
}
\label{fig:corner_plots}
\end{figure*}

\section{Results}\label{sec:results}

Applying the methodology described in Sec.~\ref{sec:methodology} to each of the trial solutions sampled for a given pulsar, we obtained one posterior value per trial solution for each of the pulsars in Table~\ref{table:input_data_table}. However, we present here only the results for a subsample of 7 MSPs that NANOGrav extensively observed. Their extensive observing baseline will serve as a benchmark for timing-based astrometry and provide a lower bound of the improvements achievable with VLBI-informed priors.

For each pulsar in Fig.~\ref{fig:corner_plots}, the heatmap in each subplot shows the posterior distribution as a function of one pair of astrometric parameters, after marginalizing (summing) the posterior values over the remaining three parameters. The posterior values have been normalized between 1 (yellow) and 0 (dark purple). We also show the NG15 astrometric estimates as red error bars. The exact maximum-posterior astrometric values for this subsample are presented and compared to the corresponding NG15 timing values in Table~\ref{tab:old_vs_new_astrometric_parameters}.


We observe in Fig.~\ref{fig:corner_plots} that the distributions are pronouncedly unimodal, with only weak correlations for a few pulsars. This indicates that the VLBI-informed priors tend to favor a single, well-defined, maximum-posterior astrometric solution per pulsar. Only using logarithmic scaling do we observe minor bimodality in the posterior distribution for the parallax. We attribute such bimodality to a covariance between the parallax and the proper motion caused by sub-optimal VLBI observing cadence and geometry.

\begin{deluxetable*}{l l r@{}r@{\fs}l r@{}r@{\farcs}l r@{}l r@{}l r@{}l}
\tablenum{2}
\tablecaption{Comparison between the timing-derived astrometric estimates reported in NG15 (after applying proper motion corrections to match the reference epochs used in the VLBI) and the maximum posterior astrometric estimates obtained in this work. The error bars for the maximum posterior estimates were approximated as half the spacing between two consecutive sampled parameter values. \label{tab:old_vs_new_astrometric_parameters}}
\tablehead{
  \colhead{PSR} & \colhead{} &
  \multicolumn{3}{c}{$\alpha$ (J2000)} &
  \multicolumn{3}{c}{$\delta$ (J2000)} &
  \multicolumn{2}{c}{$\mu_\alpha$ [mas/yr]} &
  \multicolumn{2}{c}{$\mu_\delta$ [mas/yr]} &
  \multicolumn{2}{c}{$\varpi$ [mas]}
}
\startdata
  & NG15
  & 00$^\mathrm{h}$30$^{\mathrm m}$&27&42504(3)
  & 04$\degree$51$'$&39&714(1)
  & $-$6&.26(7) & 0&.59(17) & 3&04(5) \\ J0030$+$0451
  & Max Post.
  & 00$^\mathrm{h}$30$^{\mathrm m}$ & 27&42497(40)
  & 04$\degree$51$'$ & 39&71637(17)
  & $-$6&.11(6) & 0&.22(15) & 2&92(4) \\
  & Difference
  & & \textbf{0}& \textbf{00007}
  & & \textbf{$-$0}& \textbf{00237}
  & \textbf{$-$0}& \textbf{15} & \textbf{0}& \textbf{.37} & \textbf{0}& \textbf{.12} \\
\hline
  & NG15 & 16$^\mathrm{h}$40$^\mathrm{m}$ & 16&745878(2) & 22$\degree$24' & 08&76358(4) & 2&.095(7) & $-$11&.320(7) & 0&.59(19) \\ J1640$-$2224
  & Max Post. & 16$^\mathrm{h}$40$^\mathrm{m}$ & 16&74587(4) & 22\degree24' & 08&76369(4) & 2&.109(7) & $-$11&.326(6) & 0&.60(8) \\
  & Difference & & \textbf{0}& \textbf{000008} & & \textbf{$-$0}& \textbf{00011} & \textbf{$-$0}& \textbf{.014} & \textbf{0}& \textbf{.006} & \textbf{$-$0}& \textbf{.01} \\
\hline
  & NG15 & 17$^\mathrm{h}$30$^\mathrm{m}$ & 21&67956(9) & $-$23$\degree$04' & 31&206(23) & 20&.98(78) & 15&.44(1402) & 2&.27(40)\\ J1730$-$2304
  & Max Post. & 17$^\mathrm{h}$30$^\mathrm{m}$ & 21&6797(8) & $-$23$\degree$04' & 31&17449(27) & 20&.34(4) & 3&.98(764) & 1&.82(4) \\
  & Difference & & \textbf{$-$0}& \textbf{00014} & & \textbf{0}& \textbf{03151} & \textbf{0}& \textbf{.64} & \textbf{11}& \textbf{.46} & \textbf{0}& \textbf{.45} \\
\hline
  & NG15 & 19$^\mathrm{h}$18$^\mathrm{m}$ & 48&0295004(13) & $-$06$\degree$42'& 34&9346(2) & $-$7&.146(3) & $-$5&.95(3) & 0&.76(11)\\ J1918$-$0642
  & Max Post. & 19$^\mathrm{h}$18$^\mathrm{m}$ & 48&029498(13) & $-$06$\degree$42'& 34&93(3) & $-$7&.156(2) & $-$5&.90(2) & 0&.48(6) \\
  & Difference & & \textbf{$-$0}& \textbf{0005} & & \textbf{$-$0}& \textbf{0046} & \textbf{0}& \textbf{.01} & \textbf{$-$0}& \textbf{.05} & \textbf{0}& \textbf{.28} \\
\hline
  & NG15 & 20$^\mathrm{h}$10$^\mathrm{m}$ & 45&921123(3) & $-$13$\degree$23'&56&0821(3) & 2&.56(1) & $-$5&.86(9) & 0&.51(11)\\ J2010$-$1323
  & Max Post. & 20$^\mathrm{h}$10$^\mathrm{m}$ & 45&921122(28) & $-$13$\degree$23'&56&0826(2) & 2&.56(7) & $-$5&.71(6) & 0&.55(7) \\
  & Difference & & \textbf{0}& \textbf{000001} & & \textbf{$-$0}& \textbf{0005} & \textbf{0}& \textbf{.0} & \textbf{$-$0}& \textbf{.15} & \textbf{$-$0}& \textbf{.04} \\
\hline
  & NG15 & 21$^\mathrm{h}$45$^\mathrm{m}$ & 50&45887(1) & $-$07$\degree$50'&18&5137(8) & $-$9&.55(5) & $-$8&.97(20) & 1&.59(16) \\ J2145$-$0750
  & Max Post. & 21$^\mathrm{h}$45$^\mathrm{m}$ & 50&4589(2) & $-$07$\degree$50'&18&5145(8) & $-$9&.45(5) & $-$9&.17(20) & 1&.79(4) \\
  & Difference & & \textbf{$-$0}& \textbf{00003} & & \textbf{$-$0}& \textbf{0008} & \textbf{$-$0}& \textbf{.1} & \textbf{0}& \textbf{.2} & \textbf{$-$0}& \textbf{.2} \\ \hline
  & NG15 & 23$^\mathrm{h}$17$^\mathrm{m}$ & 09&236395(1) & 14$\degree$39'&31&2653(1) & $-$1&.312(3) & 3&.45(2) & 0&.64(11) \\ J2317+1439 & Max Post. & 23$^\mathrm{h}$17$^\mathrm{m}$ & 09&23640(1) & 14$\degree$39'&31&26511(8) & $-$1&.309(3) & 3&.44(2) & 0&.52(8) \\
  & Difference & & \textbf{$-$0}& \textbf{000005} & & \textbf{0}& \textbf{00019} & \textbf{$-$0}& \textbf{.003} & \textbf{0}& \textbf{.01} & \textbf{0}& \textbf{.12} \\
\enddata
\end{deluxetable*}

For some pulsars, such as PSRs J1640$-$2224 and J1730$-$2304, the maximum-posterior estimates are largely consistent with the uncertainties of the timing estimates. However, for others like PSRs J0030$+$0451 and J2145$-$0750, the maximum posterior solution deviates by up to $2\sigma$ from the NG15 timing solution. If we assume the maximum-posterior estimates provide a more accurate astrometric approximation, these offsets could be indicative of biases in current timing-derived astrometric values.

\begin{figure*}[htp]

\centering
\subfigure[J0030$+$0451]{\includegraphics[width=0.49\linewidth]{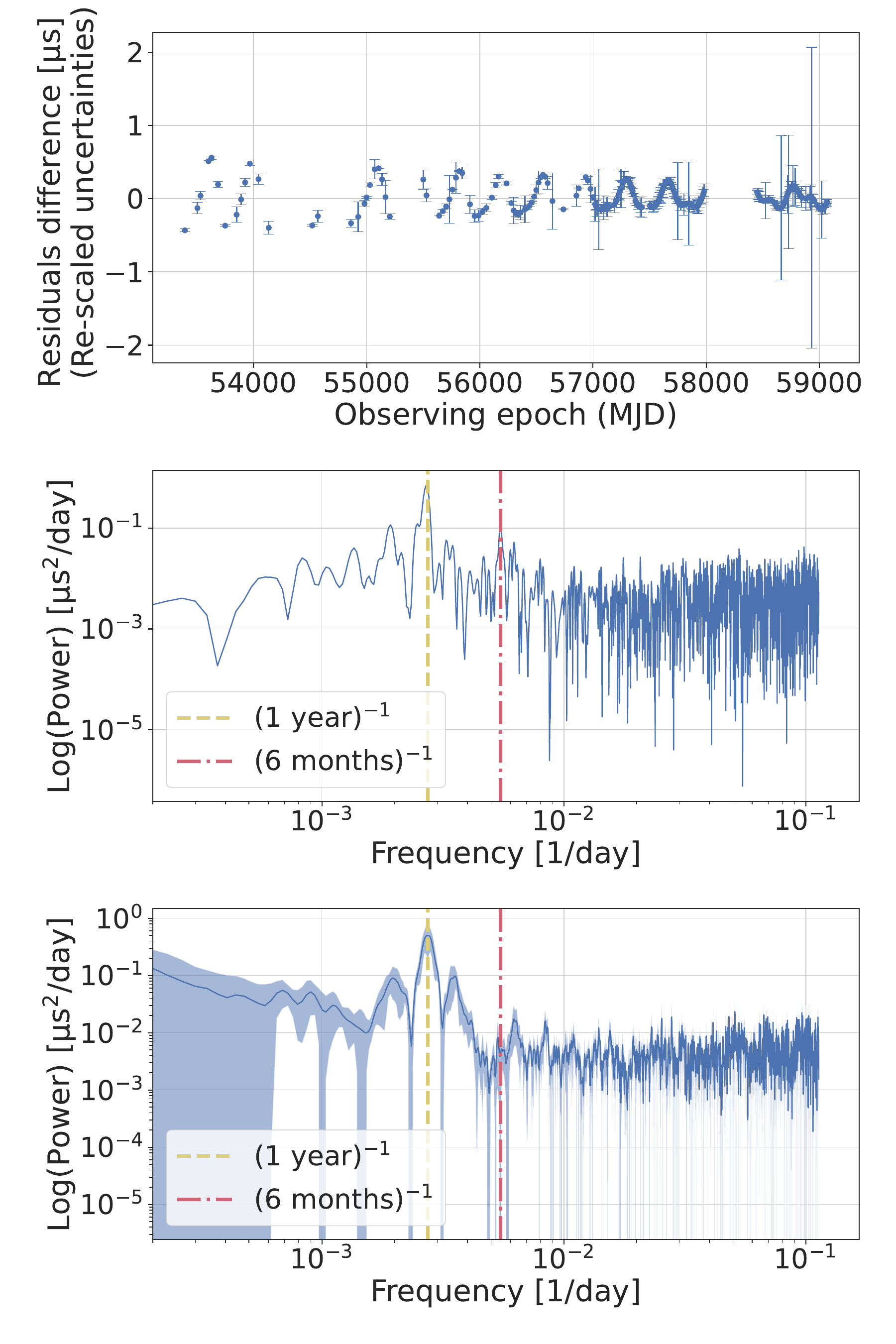}\label{fig:J0030+0451_periodgram}}
\centering
\subfigure[J2145$-$0750]{\includegraphics[width=0.49\linewidth]{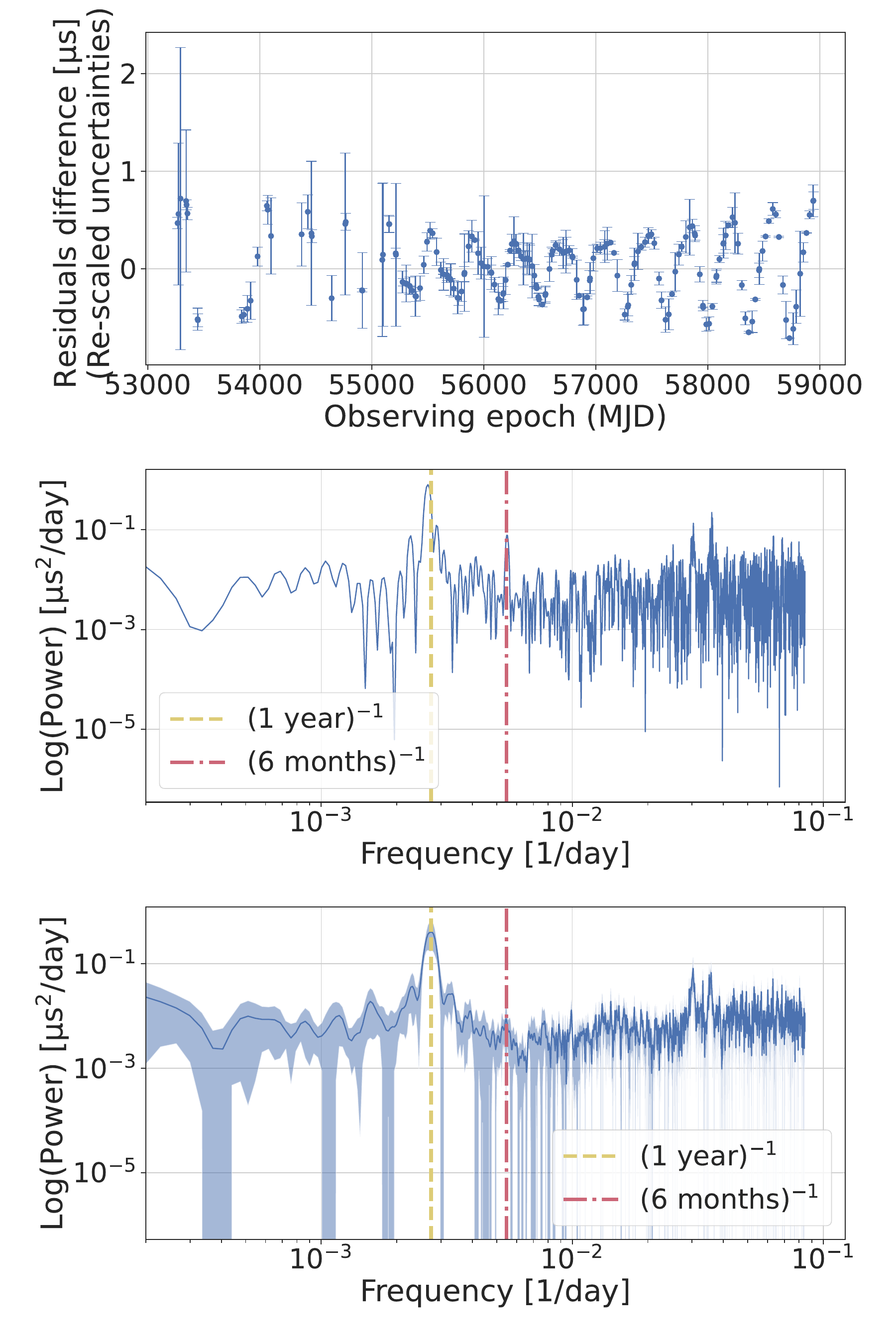}\label{fig:J2145-0750_periodogram}}

\caption{Top row: epoch- and frequency-averaged differences between the residuals obtained using either the NG15 timing model or the maximum posterior astrometric solution, based on NG15 TOA measurements for each pulsar. For visualization purposes, the error bars have been scaled by a factor $1/10$. Middle row: Lomb-Scargle periodograms of the residual differences shown in the first row. The vertical dashed lines mark the frequencies corresponding to $(1~\mathrm{year})^{-1}$ and $({6~\mathrm{months}})^{-1}$, respectively. Bottom row: The dark blue line represents the weighted average of the individual Lomb-Scargle periodograms corresponding to the residual difference for each trial timing solution using their (normalized) posteriors as the corresponding weights; the shaded blue area represents the 1$\sigma$ standard deviation at each frequency.}\label{fig:periodograms}

\end{figure*}

For each pulsar, we next compared the timing residuals obtained from the maximum-posterior solution and the NG15 timing model for each pulsar using the following procedure:

\begin{enumerate}
    \item Since in NG15 a single observing epoch can comprise multiple TOA measurements across different frequency bands, we first used PINT's \texttt{ecorr\_average} function to epoch-average the pulsar's TOAs using the ECORR noise model time-binning while also applying the noise model terms \citep[EFAC, EQUAD, ECORR; see][]{Arzoumanian_2018} to the TOA uncertainties.
    \item For each observing epoch, we performed weighted averaging of the TOAs across the frequency bands, using weights given by the square inverse of the fully-adjusted TOA uncertainty.
    \item Using the analysis scheme described in \cite{Agazie_2023b}, we fit the TOAs to two timing models: 1) NG15's timing model, and 2) the same model after replacing its astrometric values with the maximum-posterior astrometric solution. We thus obtained two sets of residuals and calculated the difference between them.
\end{enumerate}
\noindent These differences are shown for two selected pulsars in the first row of Fig.~\ref{fig:periodograms}. For visualization purposes, the error bars have been scaled by a factor $1/10$. However, a separate analysis of the residuals and their uncertainties at each epoch confirms that the differences are not consistent with zero. Instead, we observe a structure consistent with a superposition of sinusoidal components with a period of 1 year, as expected for residuals derived from timing solutions with mismatching astrometric positions.

For PSR J0030$+$0451, the amplitude of the sinusoidal structure is larger in the earlier observations, which also have smaller uncertainties, and diminishes in later observations, which exhibit larger error bars. The difference in the uncertainties can be attributed to the increased (weekly) observational cadence at Arecibo in the later part of the data set, which is weighted more heavily. However, modulation in the residual amplitude is consistent with the mismatching proper-motion values between the two timing solutions. Most importantly, the amplitude of the observed offset reaches up to $\sim0.8~\mathrm{\mu s}$, offering insight into the scale of the signals that could be absorbed into the pulsar's astrometric fit.


The residual differences between both models (top row of Fig.~\ref{fig:periodograms}) reveal periodic signals that are currently being absorbed into the astrometric fit. We can isolate and identify such signals in the residual differences using periodograms. Given the uneven cadence of the pulsar observations, we employed a Lomb-Scargle periodogram, as shown in the middle row of Fig.~\ref{fig:periodograms}. We find strong spectral power at frequencies of $\mathrm{(1~year)}^{-1}$ and $\mathrm{(6~months)}^{-1}$, as expected from differences in the pulsar position and parallax. If such biases are present in the current timing-based astrometric estimates, they would significantly reduce the detector sensitivity at these specific frequencies. This result has important implications for GW searches targeting sources whose emission frequencies coincide with these peaks in power absorption.

The lower frequency range we can analyze is limited by the finite time baseline of the timing data set ($~$15 years). However, in both periodograms, we notice generally flat power spectra at Fourier frequencies $\lesssim10^{-3}$, with no excess of spectral power or sharp features that could be attributed to power contamination due to astrometry-induced signals. This suggests that, for these pulsars and assuming that the maximum posterior solution represents the best approximation to the real astrometry, the astrometric fit would not absorb a significant amount of red noise power at lower frequencies, such as that associated with a stochastic GW background.

Nevertheless, this periodogram only shows the power absorption expected under the assumption that the highest-posterior astrometric solution represents a better approximation to the real astrometry. Other astrometric solutions may still have non-negligible probabilities of representing better approximations. To analyze the behavior over the broader parameter space, we repeated the calculation of a Lomb-Scargle periodogram for every trial astrometric solution. We then took the weighted average of the spectral power at each frequency across the individual periodograms using the normalized posterior of each trial solution as the corresponding weight. The results for PSRs J0030$+$0451 and J2145$-$0750 are presented in the bottom row of Fig.~\ref{fig:periodograms}.

On the sides of the peak at $\mathrm{(1~year)}^{-1}$ observed in the previous periodogram, we now see two side lobes which do not correspond to real physical processes but are rather an artifact from the finite 15-year time baseline of our observations \citep[see Fig. 4d in][]{hrs2019}. More importantly, we also find a steeper power-law index at lower frequencies. Therefore, as we depart from the maximum-posterior astrometric solution, we find significant absorption of spectral power corresponding to red noise into the astrometric fit.

\section{Conclusions}\label{sec:conclusions}

In this work, we set out to investigate (1) whether VLBI-derived measurements of pulsar astrometric parameters can be effectively used as priors when calculating timing solutions, and (2) what the consequences of incorporating such priors are for timing precision and noise absorption.

To address the first question, we combined the frame tie formalism introduced by \cite{2013ApJ...777..104M} and later refined by \cite{2017MNRAS.469..425W} with high-precision astrometric measurements from projects like MSPSR$\pi$ \citep{2023MNRAS.519.4982D}. In doing so, we derived a frame tie between the ICRS (as realized by the RFC) and the timing reference system defined by the JPL DE440 Solar System ephemeris. Applying this frame tie to convert VLBI astrometric estimates into the timing reference system resulted in an overall improved agreement between interferometric and timing measurements, compared to using uncorrected coordinates. However, we note two key limitations in this process:

\begin{itemize}
    \item When calculating the covariance matrix for the interferometric positions, it was necessary to neglect potential correlations between right ascension and declination measurements, as these are often omitted from published results. As noted by \cite{2017MNRAS.469..425W}, we encourage VLBI observers to include these correlations in future publications.
    \item In Sec.~\ref{sec:vlbi_observations}, we addressed systematic offsets between different VLBI campaigns caused by the use of primary phase calibrators from different catalogs, resulting in different realizations of the ICRS. However, VLBI measurements also rely on secondary phase calibrators (fainter sources located closer to the target field) to refine phase calibration for localized atmospheric effects. Future analyses should consider the positional offsets of these secondary calibrators across catalogs and their potential impact on the resulting astrometric measurements.
\end{itemize}

With this frame tie correction in place, we used VLBI-derived astrometric estimates to create priors when fitting timing solutions for 18 MSPs included in the NG15 dataset. We found that the resulting maximum-posterior astrometric solutions can deviate from the corresponding NG15 timing values by up to $\sim 2 \sigma$. If we interpret the maximum posterior solution as a higher-accuracy astrometric estimate, this offset suggests that NG15's astrometric values, while precise, may suffer from biases.

Such biases in the timing astrometry could arise from red noise power being partially absorbed into astrometric parameter fitting, thereby distorting the resulting parameter estimates and reducing sensitivity to red processes, including GWs. To quantify this effect, we compared timing residuals calculated using NG15 TOAs and either the NG15 or maximum-posterior timing models. We observed sinusoidal structures in the residual differences, with amplitudes of up to $\sim0.8~\mathrm{\mu s}$ (see Fig.~\ref{fig:periodograms}). This offset provides a scale of the residual amplitudes that could be absorbed in the astrometric fit. In particular, for PSR J0030$+$0451, we found that the amplitude of this offset is more pronounced in earlier observations, which also have smaller uncertainties, suggesting the presence of error mis-estimations in earlier data, such as those examined in \cite{Sosa_Fiscella_2024}.

In analyzing the periodogram resulting from these residual differences, we found that, for the pulsars analyzed in this work and under the assumption that the maximum posterior solution represents the best estimate to the real astrometry, the astrometric offsets in NANOGrav's timing model would not lead to a significant absorption of red noise power at lower frequencies, such as that associated with a stochastic GW background. However, as we depart from this maximum-posterior solution, we do find significant power absorption into the astrometric fit.

We also found significant power absorption at frequencies of $\mathrm{(1~year)}^{-1}$ and $\mathrm{(6~months)}^{-1}$. These results are of particular interest for GW searches from sources emitting at these frequencies. For example, the radio galaxy 3C 66B has been hypothesized to host a supermassive black hole binary at its center with an orbital period of $1.05 \pm 0.03$ years based on electromagnetic observations \citep{doi:10.1126/science.1082817}. Given the proximity of this source to our Galaxy, the estimated binary parameters imply a GW amplitude that should have been well within the detection capabilities of pulsar timing experiments \citep{Agazie_2024}. However, previous timing-based targeted searches 
\citep[e.g.,][]{Agazie_2024, 2025arXiv250814742T, 2025arXiv250820007C} have not detected a significant GW signal consistent with the proposed orbital period. Given our findings, a possible explanation for this absence of detections is that the residual amplitude introduced by a GW signature in timing observations is being mistakenly attributed to errors in the pulsar position, resulting in a GW power absorption into the astrometric fit. In turn, not only would this effect limit the GW sensitivity, but also bias the astrometric estimates derived from timing.

In conclusion, VLBI provides independent and high-precision astrometric measurements that, when incorporated into timing models via methods like outs, can directly reduce the number of free parameters in timing fits. This reduction would mitigate astrometric biases and enhance sensitivity to red processes, including gravitational waves. VLBI-informed priors are particularly valuable for newly discovered pulsars whose timing baselines are too short to yield reliable astrometric solutions. With adequate VLBI data and a properly calibrated frame tie, VLBI can constrain positions, proper motions, and parallaxes to milliarcsecond precision far earlier than with timing alone.

Once a long enough observing baseline has been established and if the timing RMS falls below several microseconds, timing can yield more precise astrometric estimates than VLBI \citep{2013ApJ...777..104M}. However, for pulsars with substantial timing noise, even long baselines can yield biased astrometric fits. Therefore, VLBI will remain essential both in better constraining the astrometric parameters of both newly discovered pulsars and in well-observed pulsars with high levels of noise.

Looking ahead, future VLBI campaigns should include a broader sample of MSPs and, crucially, also report the correlation between their right ascension and declination measurements. Such datasets could greatly help refine the frame tie between the VLBI and timing reference frames. In turn, this refined frame tie would not only yield improved VLBI-based astrometric priors, but it is also of astrometric interest for expanding our understanding of the relationship between different reference frames.


\section*{Acknowledgments}

S.V.S.F. undertook the analysis, developed the code pipeline, and prepared the ﬁgures, tables, and the majority of the text. M.T.L. developed the mathematical framework for this work, selected the analyzed data set, assisted with the preparation of the manuscript, provided advice on interpreting the results, and supervised the project development. J.S.H. and M.S.D.B. assisted with the preparation of the manuscript. With the exception of S.V.S.F. and M.S.D.B., all other authors were involved in curating the NANOGrav 15-year data set; specific contributions are summarized in \cite{NG15}.

S.V.S.F. is supported by the National Science Foundation Graduate Research Fellowship Program under Grant No. 2139292, and acknowledges partial support from the NASA New York Space Grant. P.R.B.\ is supported by the Science and Techmarginalizedy Facilities Council, grant number ST/W000946/1. H.T.C.\ acknowledges funding from the U.S. Naval Research Laboratory. Pulsar research at UBC is supported by an NSERC Discovery Grant and by CIFAR. K.C.\ is supported by a UBC Four Year Fellowship (6456). M.E.D.\ acknowledges support from the Naval Research Laboratory by NASA under contract S-15633Y. T.D.\ and M.T.L.\ received support by an NSF Astronomy and Astrophysics Grant (AAG) award number 2009468 during this work. E.C.F.\ is supported by NASA under award number 80GSFC24M0006. G.E.F.\ is supported by NSF award PHY-2011772. D.C.G.\ is supported by NSF Astronomy and Astrophysics Grant (AAG) award \#2406919. D.R.L.\ and M.A.M.\ are supported by NSF \#1458952. M.A.M.\ is supported by NSF \#2009425. The Dunlap Institute is funded by an endowment established by the David Dunlap family and the University of Toronto. T.T.P.\ acknowledges support from the Extragalactic Astrophysics Research Group at E\"{o}tv\"{o}s Lor\'{a}nd University, funded by the E\"{o}tv\"{o}s Lor\'{a}nd Research Network (ELKH), which was used during the development of this research. H.A.R.\ is supported by NSF Partnerships for Research and Education in Physics (PREP) award No.\ 2216793. S.M.R.\ and I.H.S.\ are CIFAR Fellows. Portions of this work performed at NRL were supported by ONR 6.1 basic research funding.

We acknowledge support from the NSF Physics Frontiers Center award number 2020265, which supports the NANOGrav project. The authors acknowledge Research Computing at the Rochester Institute of Techmarginalizedy for providing computational resources and support that have contributed to the research results reported in this publication. The Green Bank Observatory and National Radio Astronomy Observatory are facilities of the NSF operated under a cooperative agreement by Associated Universities, Inc.



This work made use of Astropy:\footnote{http://www.astropy.org} a community-developed core Python package and an ecosystem of tools and resources for astronomy \citep{astropy:2013, astropy:2018, astropy:2022}. 


%

\vspace{5mm}
\facilities{Green Bank Observatory (GBO), Arecibo Observatory (AO), Very Large Array (VLA), Very Long Baseline Array (VLBA).}


\software{PINT \citep{Luo_2021}, Astropy \citep{astropy:2013, astropy:2018, astropy:2022}.}



\appendix

\section{Reproducing our work}
\label{sec:appendix}

Our software scripts, input data files, and the order in which they must be executed are available from our GitHub repository (\url{https://github.com/sophiasosafiscella/VLBI_timing}). Our Python code requires the \texttt{astropy}, \texttt{PINT}, and \texttt{uncertainties} packages. The input \texttt{.par} and \texttt{.tim} pulsar timing data files are available from \url{https://zenodo.org/records/17188627}. The timing astrometric data were processed using \texttt{PINT}, and the results were recorded in the file \texttt{timing\_astrometric\_data\_updated.csv}. The VLBI positions are obtained from the literature and stored in \texttt{msp\_vlbi.csv}. The calibration source positions for the millisecond pulsars are stored in \texttt{cal.csv}.

\section{Linearized Catalogue-to-Catalogue Transformation}
\label{sec:linearized}

The difference between two realizations of the ICRS materialized through different VLBI catalogs is, fundamentally, a three-dimensional rotation. However, for phase-referenced VLBI, the relevant correction is local. Linearizing the rotation shows that applying the calibrator’s RFC--catalog offset to the pulsar reproduces the true rotated pulsar position to first order.  The neglected term is of order $\mathcal{O}(|\boldsymbol{\epsilon}|\,s)$, and is negligible when both the rotation $|\boldsymbol{\epsilon}|$ and the separation $s$ are small. In this appendix, we will derive this residual error and prove that it is negligible for the precision levels involved in this work.

Let $\widehat{\mathbf{n}}^{\rm cat}$ be a unit vector pointing to a source in a given catalogue frame (i.e., ICRF1, NBSS, VCS1, etc.) and $\widehat{\mathbf{n}}^{\rm RFC}$ the same source in the RFC frame. The two frames differ by a small rotation generated by the vector $\boldsymbol{\epsilon}$ (in radians), so the exact relation is
\begin{equation}
    \widehat{\mathbf{n}}^{\rm RFC} = \boldsymbol{\Omega} \, \widehat{\mathbf{n}}^{\rm cat} = \exp\!\big([\boldsymbol{\epsilon}]_\times\big)\,\widehat{\mathbf{n}}^{\rm cat},
\end{equation}

\noindent where $[\boldsymbol{\epsilon}]_\times$ is the skew matrix of the vector $\boldsymbol{\epsilon}$. For $|\boldsymbol{\epsilon}|\ll1$ we can linearized:

\begin{equation}
    \widehat{\mathbf{n}}^{\rm RFC} \approx 
    \big(\mathbf{I} + [\boldsymbol{\epsilon}]_\times\big)\,\widehat{\mathbf{n}}^{\rm cat}
    = \widehat{\mathbf{n}}^{\rm cat} + \boldsymbol{\epsilon}\times\widehat{\mathbf{n}}^{\rm cat}.
    \label{eq:linearization}
\end{equation}

If we define the calibrator and pulsar unit vectors in the catalogue frame as 
$\widehat{\mathbf{n}}^{\rm cat}_{\rm cal}$ and 
$\widehat{\mathbf{n}}^{\rm cat}_{\rm psr}$, and their RFC counterparts similarly, then from Eq.~\ref{eq:linearization} we see that the observed difference for the calibrator is
\begin{equation}
    \Delta\widehat{\mathbf{n}}_{\rm cal} \equiv
    \widehat{\mathbf{n}}^{\rm RFC}_{\rm cal} - \widehat{\mathbf{n}}^{\rm cat}_{\rm cal}
    \approx \boldsymbol{\epsilon}\times\widehat{\mathbf{n}}^{\rm cat}_{\rm cal}.
    \label{eq:calib_offset}
\end{equation}

If we apply the calibrator offset directly to the pulsar, the linear correction gives
\begin{equation}
    \widehat{\mathbf{n}}^{\rm cat}_{\rm psr} \;\mapsto\;
    \widehat{\mathbf{n}}^{\rm cat}_{\rm psr} + \Delta\widehat{\mathbf{n}}_{\rm cal}.
\end{equation}

\noindent The true rotated pulsar vector, however, is
\begin{equation}
    \widehat{\mathbf{n}}^{\rm RFC}_{\rm psr}
    \approx \widehat{\mathbf{n}}^{\rm cat}_{\rm psr} + \boldsymbol{\epsilon}\times\widehat{\mathbf{n}}^{\rm cat}_{\rm psr}.
\end{equation}

\noindent Subtracting these two expressions gives the residual error:
\begin{equation}
    \boldsymbol{\Delta}_{\rm res}
    \equiv \widehat{\mathbf{n}}^{\rm RFC}_{\rm psr} - \Big(\widehat{\mathbf{n}}^{\rm cat}_{\rm psr} + \Delta\widehat{\mathbf{n}}_{\rm cal}\Big)
    \approx \boldsymbol{\epsilon}\times\big(\widehat{\mathbf{n}}^{\rm cat}_{\rm psr} - \widehat{\mathbf{n}}^{\rm cat}_{\rm cal}\big).
    \label{eq:residual}
\end{equation}

Let $s$ be the angular separation between pulsar and calibrator. Eq.~\eqref{eq:residual} implies
\begin{equation}
    \|\boldsymbol{\Delta}_{\rm res}\| \sim |\boldsymbol{\epsilon}|\,s.
\end{equation}
Thus, the error introduced by using the linear ``apply calibrator offset'' prescription is the product of the global frame-tie angle $|\boldsymbol{\epsilon}|$ and the pulsar--calibrator separation $s$. For typical values (frame differences at the $\mu$as--mas level and separations of a few degrees), this residual is negligibly small compared with current VLBI astrometric uncertainties.


\bibliographystyle{mnras}
\bibliography{main}



\end{document}